\documentclass[preprint2]{aastex}

\shorttitle{Nonthermal radiation of young SNRs} \shortauthors{V.N.
Zirakashvili, F.Aharonian}

\begin{document}


\title{NONTHERMAL RADIATION OF YOUNG SUPERNOVA REMNANTS}


\author{V.N.Zirakashvili}
\affil{Pushkov Institute for Terrestrial Magnetism, Ionosphere and Radiowave
Propagation, 142190, Troitsk, Moscow Region, Russia}
\affil{Max-Planck-Institut f\"{u}r\ Kernphysik, Saupfercheckweg 1, 69117
Heidelberg, Germany}
\author{F.A.Aharonian}
\affil{Dublin Institute for Advanced Studies, 31 Fitzwilliam Place, Dublin 2, Ireland}
\affil{Max-Planck-Institut f\"{u}r\ Kernphysik,  Saupfercheckweg 1, 69117
Heidelberg, Germany}



\begin{abstract}
A new numerical code, designed for the detailed numerical treatment of
nonlinear diffusive shock acceleration, is used for modeling of particle
acceleration and radiation in young supernova remnants. The model is based
on spherically symmetric hydrodynamic equations complemented with transport
equations for relativistic particles. For the first time, the acceleration
of electrons and protons by both forward and reverse shocks is studied
through detailed numerical calculations. We model the energy spectra and
spatial distributions of nonthermal emission of the young supernova remnant
RX J1713.7-3946 and compare the calculations with the spectral and
morphological properties of this object obtained in broad energy band from
radio to very high energy gamma-rays. We discuss the advantages and
shortcomings of the so-called hadronic and leptonic models which assume that
the observed TeV gamma-ray emission is produced by accelerated protons and
electrons, respectively. We discuss also a "composite" scenario when the
gamma-ray flux from the main parts of the shell has inverse Compton
origin, but with a non-negligible contribution of hadronic origin from dense
clouds interacting with the shell.

\end{abstract}





\section{Introduction}

The  paradigm of the diffusive shock acceleration (DSA) of relativistic particles
 (Krymsky \citealp{krymsky77}, Axford et al. \citealp{axford77}, Bell
\citealp{bell78}, Blandford \& Ostriker \citealp{blandford78})  is generally accepted
as the  most likely scenario  of production of
galactic cosmic rays (CR) in supernova remnants (SNRs). Over the last 30 years
a significant progress has been achieved in the development of theoretical models and
understanding  of  the  basic features of DSA   (see e.g. Malkov \& Drury
\citealp{malkov01} for a review).  On the other hand, the recent detailed studies
of spectral and morphological features of young SNRs, first of all in the
X-ray and very-high-energy or TeV gamma-ray band, provide excellent observational
material for  development of detailed  numerical models of acceleration and radiation
of relativistic electrons and protons in young SNRs.  These observations generally confirm
in general terms the  predictions of DSA. In particular  the synchrotron X-radiation observed from  many young
SNRs implies an existence of multi-TeV energy electrons which is naturally explained by DSA.
The detection of TeV gamma-rays from SNRs, like Cas~A, RX J1713.7-3946, Vela Jr.,
RCW~86 (see Aharonian et al. \citealp{ROP} for a recent review) give a more direct and unambiguous
information about the effective acceleration of particles, electrons and/or protons, in SNRs to energies
exceeding 100 TeV.  In spite of  the high quality data obtained  from these  SNRs and
intensive theoretical and phenomenological studies,
we cannot yet  firmly distinguish  between two competing processes of  gamma-ray
production: the  inverse Compton (IC) scattering of  the background
electromagnetic radiation by accelerated electrons  (leptonic
model) or a decay of neutral pions, mainly produced by accelerating
protons during the interaction with the gas of the remnant
(hadronic model).  It should be noted that  this ambiguity  concerns the
interpretation of the origin of gamma-rays, rather than the
question of acceleration of protons in general.   The presence of
multi-TeV electrons in the SNR shells most likely implies also acceleration of protons -
not only because the DSA  operates identically  for  both electrons and protons;
the presence of larger amount of nonthermal protons is needed in any case
for production of magnetohydrodynamic (MHD) turbulence in the vicinity of a
supernova shock. This is a necessary  condition for effective  acceleration of
particles via DSA (see  Malkov \& Drury \citealp{malkov01} for a review).

In this paper we  conduct detailed study acceleration of electrons
and protons with an emphasis on the  spectral and morphological
features of high energy radiation produced by these particles in
young supernova remnants.  For that purpose we use a new numerical
code of nonlinear diffusive shock acceleration developed by one of
us in collaboration with V. Ptuskin (Zirakashvili \& Ptuskin
\citealp{zirakashvili09}). This model can be considered as a
natural development of existing numerical codes (see e.g. Berezhko
et al.\citealp{berezhko94}, Kang et al \citealp{kang06}),  with
new additional elements  which despite their  strong impact on the
overall picture of acceleration in general, and on the properties
of  high energy radiation of SNRs in particular, have been ignored
in the past.   Namely, in our  treatment, the solution of
spherically symmetric hydrodynamic equations is combined with the
energetic particle transport and acceleration by the {\it forward}
shock (FS) and {\it reverse}  shock (RS). Nonlinear response of
energetic particles via their pressure gradient results in
self-regulation of acceleration efficiency.  To our knowledge,
this is the first study in which the modification of the reverse
shock and the nonthermal radiation  related to the reverse shock
is  taken into account. The   detailed calculations  of radio,
X-ray, gamma-ray emission components conducted within  a
self-consistent treatment of particle acceleration by both forward
and reverse shocks  allows direct comparison of the observed
spectral and morphological features with the model predictions.
The inclusion of the radiation components related to the reverse
shock seems to us  a rather obligatory condition,  given the
effective acceleration of particles at the reverse shock to very
high energies, as it is demonstrated in this paper.  In this
regard we note that  the parameters  characterizing the reverse
shock can be significantly different in comparison with the
parameters at the forward shock. As a result, the properties of
radiation components from the reverse and forward shocks can  be
also significantly  different. In particular the  density of
plasma and the magnetic field in the reverse shock can be  very
low which would dramatically increase the contribution of the IC
component compared to the hadronic ($\pi^0$-decay) component  of
gamma-rays.   Because of the higher  gas density and stronger
magnetic field, in the forward shock one may expect an opposite
relation between the contributions of the electrons  and  protons
to the high energy gamma-radiation.

The results of this study  have a general character and can be applied to different
young SNRs.   In order to demonstrate their astrophysical implication, in this paper we apply  the model
to the supernova remnant RX J1713.7-3946 - the brightest and best studied
TeV gamma-ray source among known shell-type supernova remnants (Aharonian et al. \citealp{aharonian07}).
The high quality gamma-ray and  X-ray  images and energy spectra
available  for this source make it a kind of template source  for theoretical  studies.
On the other hand,  this source  have quite unusual  radiation features (e.g. a lack of thermal
X-radiation, a very low intensity of  the synchrotron radio emission, etc.)
so the conclusions derived  for this source cannot be directly extended to other SNRs.

The paper is organized as follows. In Section 2 we briefly summarize the observational
information about the supernova remnant RX J1713.7-3946. The short description of the
model and basic features of accelerated particles is given in Section  3. The results of
modeling of the broad-band emission radiation produced in  hadronic and leptonic scenarios
are presented  and discussed in Sections  4, 5 and 6.  In Section 7 we compare  the magnetic field
amplification and the  maximum particle energies achieved in forward and reverse shocks. Finally,
in Sections 8 and 9 we discuss and summarize the obtained results.

\section{Observational properties of RX J1713.7-3946}

The shell-type SNR RX J1713.7-3946 was discovered in X-rays during
the ROSAT All-Sky Survey (Pfeffermann \& Aschenbach
\citealp{pfeffermann96}). The following observations of ASCA,
Chandra, XMM-Newton and Suzaku   (Koyama et al.
\citealp{koyama97}, Uchiyama et al. \citealp{uchiyama03},
Cassam-Chena\"i et al. \citealp{cassam04}, Hiraga et al.
\citealp{hiraga05}, Takahashi et al. \citealp{takahashi08}, Tanaka
et al. \citealp{tanaka08}) revealed that the X-ray emission
predominantly consists of the synchrotron  component without an
indication of a thermal X-ray component at the level of
sensitivity of these instruments. The Suzaku measurements (Tanaka
et al. \citealp{tanaka08}) allowed accurate derivation of the
broad-band X-ray spectrum of the source  over two energy decades,
from 0.4 keV to 40 keV. It is described by a power-law with a
photon index  $\sim 2$ and a smooth cut-off around a few  keV.

The remnant is close to the position of the guest star AD393
described in the ancient Chinese records (Wang et al.\citealp{wang97}).
If this not an accidental coincidence,  the age of the remnant should be  close to 1600 years.

Using the association of the supernova remnant  with nearby molecular clouds,
different estimates of the distance to the source have been suggested:
$D=6$ kpc (Slane et al. \citealp{slane99}),  $D=1$ kpc (Fukui et al. \citealp{fukui03})
and $1.3\pm 0.4$ kpc (Cassam-Chena\"i et al. \citealp{cassam04}).

The remnant was claimed to be detected in gamma-rays by the  CANGAROO
collaboration (Muraishi et al. \citealp{muraishi00}).  Later, the observation with the
HESS array of imaging atmospheric Cherenkov  telescopes
revealed indeed a bright TeV gamma-ray source with a shell-type morphology
quite  similar to the synchrotron X-ray image of the source  (Aharonian et al. \citealp{aharonian04})
The  energy spectrum measured over the interval from 0.3 TeV to 100 TeV is hard; it is characterized by a
photon index $\simeq 2$ with a break  or a cutoff around
10 TeV (Aharonian et al. \citealp{aharonian07}).

The remnant is rather faint in radio. The recent ATCA measurements at
1.4 GHz gave the radio-intensity $22\pm 2$ Jy from the whole
remnant (Acero et al. \citealp{acero09}). The radio and X-ray images
clearly show also an inner shell (Ellison et al. \citealp{ellison01}, Lazendic et
al. \citealp{lazendic04}, Cassam-Chena\"i et al. \citealp{cassam04},
Hiraga et al. \citealp{hiraga05}, Acero et al. \citealp{acero09}) with an angular diameter 0.5 degrees
while the  angular diameter of the remnant is close to one degree.
The inner shell in the radio and X-ray images can be naturally attributed to the reverse shock
propagating in the supernova ejecta, assuming  that the electrons are accelerated also at the
reverse shock.   This is likely  to be the case of the supernova remnant Cas A (Helder \& Vink \citealp{helder08}).
Generally, the reverse shock is
not treated  as an efficient accelerator, because the magnetic field of ejecta might be very weak
due to the large  expansion factor of the exploded star. However, similar to the case of the forward shock,
the  magnetic field can be  significantly amplified  also at the reverse shock (see Ellison et al. \citealp{ellison05}), for example,
in the course of the nonresonant streaming instability introduced by Bell  \citealp{bell04}.

Another probable explanation of the inner shell could be its  association
with a pulsar wind nebula. It is known that an active pulsar may
produce a corresponding nebula in an expanding supernova ejecta
(Chevalier \citealp{chevalier05}). The inner shell does contain the
radio pulsar PSR J1713-3945 with a period $P=392$ ms, but  the latter
is located at a distance $D=4.3$ kpc.  Also, the pulsar's
current spin-down luminosity  does not exceed  $10^{34}$ erg s$^{-1}$.
This power is too low  to produce the inner shell.The shell contains also
a central compact object 1WGA J1713.4-3949
which  is presumably a  neutron star related to  the remnant
(Lazendic et al. \citealp{lazendic03}, Cassam-Chena\"i et al. \citealp{cassam04})

\section{Nonlinear model of diffusive shock acceleration}

Hydrodynamical equations for the gas density  $\rho (r,t)$, gas velocity $u(r,t)$, gas pressure
$P_g(r,t)$, and the equation for the quasi-isotropic CR momentum distribution
 $N(r,t,p)$ in the spherically symmetrical  case are given by

\begin{equation}
\frac {\partial \rho }{\partial t}=-\frac {1}{r^2}\frac {\partial }{\partial r}r^2u\rho
\end{equation}

\begin{equation}
\frac {\partial u}{\partial t}=-u\frac {\partial u}{\partial r}-\frac {1}{\rho }
\left( \frac {\partial P_g}{\partial r}+\frac {\partial P_c}{\partial r}\right)
\end{equation}

\begin{equation}
\frac {\partial P_g}{\partial t}=-u\frac {\partial P_g}{\partial r}
-\frac {\gamma _gP_g}{r^2}\frac {\partial r^2u}{\partial r}
-(\gamma _g-1)(w-u)\frac {\partial P_c}{\partial r}
\end{equation}

\[
\frac {\partial N}{\partial t}=\frac {1}{r^2}\frac {\partial }{\partial r}r^2D(p,r,t)
\frac {\partial N}{\partial r}
-w\frac {\partial N}{\partial r}+\frac {\partial N}{\partial p}
\frac {p}{3r^2}\frac {\partial r^2w}{\partial r}
\]
\[
+\frac {\eta _f\delta (p-p_{f})}{4\pi p^2_{f}m}\rho (R_f+0,t)(\dot{R}_f-u(R+0,t))\delta (r-R_f(t))
\]
\begin{equation}
+\frac {\eta _b\delta (p-p_{b})}{4\pi p^2_{b}m}\rho (R_b-0,t)(u(R_b-0,t)-\dot{R}_b)\delta (r-R_b(t))
\end{equation}
Here $P_c=4\pi \int p^2dpvpN/3$ is the CR pressure, $w(r,t)$ is the  advective velocity of CRs,
$\gamma _g$ is the adiabatic index of the gas and $D(r,t,p)$ is the CR diffusion coefficient.
It was assumed that the diffusive streaming of CRs results in the generation of magnetohydrodynamic (MHD)
waves. CR particles are scattered by these waves. That is why the CR advective velocity
 $w$ may differ from the gas velocity $u$. Damping of these waves results in an additional gas heating. It is
described by the last term in Eq. (3). Two last terms in Eq. (4)
correspond to the injection of thermal protons with momenta
$p=p_{f}$, $p=p_{b}$ and mass $m$ at the fronts of the forward and
reverse shocks at $r=R_f(t)$ and $r=R_b(t)$
respectively. The dimensionless parameters $\eta _f$ and $\eta _b$ determine
the corresponding injection efficiency\footnote{Throughout the paper we use
the  subscripts "f"  and "b"  to  the parameters characterizing the forward and reverse
(backward) shocks, respectively.}.

The pressure of energetic electrons is neglected, i.e.
the electrons are treated as test particles. The evolution of
electron distribution is described by an equation similar to Eq.(4),
but  with additional terms describing synchrotron and IC losses.

CR diffusion is determined by magnetic inhomogeneities. Strong
streaming of accelerated particles changes medium properties in
the shock vicinity. In particular, the CR streaming instability in young SNRs
 results in a high level of MHD turbulence  (Bell \citealp{bell78})
and even in  amplification of magnetic field  (Bell \citealp{bell04}). Due
to this effect the protons can be accelerated to energies beyond the
the so-called  Lagage and Cesarsky limit, $E \leq  100$~TeV (Lagage \& Cesarsky \citealp{lagage83}).

According to the recent numerical modeling of this instability,
the magnetic field is amplified by the flux of run-away highest energy
particles in the relatively broad region upstream of the shock (Zirakashvili \& Ptuskin
\citealp{zirakashvili08}). Magnetic energy density is a small fraction
($\sim 10^{-3}$) of the energy density of accelerated particles.
This amplified almost isotropic magnetic field can be considered
as a large-scale magnetic field for lower energy particles which
are concentrated in the narrow region upstream of the shock.
Streaming instability of these particles produces MHD waves
propagating in the direction opposite to the CR gradient. This
gradient is negative upstream of the  forward shock and MHD waves propagate
in the positive direction.

We apply a finite-difference method to solve numerically Eqs. (1-4)
upstream and downstream of the forward and reverse shocks. The
auto-model variables  $\xi
_1=r/R_f(t)$ and $\xi _4=r/R_b(t)$ are used instead of the radius $r$
upstream of the forward shock at $r>R_f$ and upstream of the
reverse shock at $r<R_b$ respectively. The gases compressed at
forward and reverse shocks are separated by a contact
discontinuity (CD) at $r=R_c$ that is situated between the shocks. We
use variables $\xi _2=(r-R_c)/(R_f-R_c)$ and $\xi
_3=(r-R_c)/(R_c-R_b)$ instead of $r$ between the shocks downstream
of the forward and reverse shocks, respectively.

A non-uniform numerical grid upstream of
the shocks at
 $r>R_f$ and $r<R_b$ allows to resolve small scales of hydrodynamical quantities appearing due to
the pressure gradient of low-energy CRs. Eq.
 (4) for CRs was solved using an implicit finite-difference scheme. An explicit conservative TVD scheme
(Trac \& Pen \citealp{trac03}) for hydrodynamical equations (1-3) and uniform spatial grid were used between the shocks.
These
equations are solved in the upstream regions  using an implicit
finite-difference scheme.

The magnetic field plays no dynamical role in the model. Since we
do not model  the amplification and transport
of magnetic field here, its coordinate dependence should be
specified for determination of cosmic ray diffusion and for
calculation of synchrotron emission and losses. We shall assume
below that the coordinate dependencies of the magnetic field and
the gas density coincide upstream and downstream of the forward
shock:
\begin{equation}
B(r)=\sqrt{4\pi \rho _0}\frac {\dot{R}_f\rho }{M^f_A\rho _0}, \ r>R_c.
\end{equation}
Here $\rho _0$ is the gas density of the circumstellar medium. The parameter
  $M^f_A$ is similar to the Alfv\'en Much number of the shock and
determines the value of the amplified magnetic field strength. The
magnetic energy downstream of the shock is estimated to be several percents of
the dynamical pressure $\rho _0\dot{R}^2$ as was derived from the
width of X-ray filaments in young SNRs (V\"olk et al.
\citealp{voelk05}). The characteristic range of this parameter is
$M^f_A\sim 10\div 40$. For example, the energy of the magnetic
field of about 3.5 percent  of the dynamical pressure  (V\"olk et
al. \citealp{voelk05}) and characteristic compression ratio of a
modified shock
 $\sigma =6$ correspond to  $M^f_A\approx 23$. Since the plasma density $\rho $ decreases
towards the contact discontinuity downstream of the forward shock,
the same is true for the magnetic field strength according to Eq. (5). This
seems reasonable because of a possible magnetic dissipation in this region.

Situation is different downstream of the reverse shock at
$R_b<r<R_c$. The plasma flow is as a rule strongly influenced by
the Rayleigh-Taylor instability that occurs in the vicinity of the
contact discontinuity and results in generation of MHD
turbulence in this region. We assume that the magnetic field
does not depend on radius downstream of the reverse shock while
the dependence in the upstream region is described by the equation
similar to Eq. (5):

\[
B(r)=\sqrt{4\pi \rho _m}\frac {|\dot{R}_b-u(r_m)|}{M^{b}_A}\times
\]
\begin{equation}
\left\{ \begin{array}{lll}
1, \ r<r_m,
\\
\rho /\rho _m, \ r_m<r<R_b,
\\
\rho (R_b+0)/\rho _m, \ R_b<r<R_c
\end{array} \right.
\end{equation}
Here $r_m<R_b$ is the radius where the ejecta density has a
minimum and equals $\rho _m$. This radius $r_m$ is generally close
to the reverse shock radius $R_b$ and is equal to it if the
reverse shock is not modified by the cosmic ray pressure.

CR advective velocity may differ from the gas velocity on the value of the radial
component of the Alfv\'en velocity
$V_{Ar}$ calculated in the isotropic random magnetic field:  $w=u+\xi _AV_{Ar}$. Here the factor $\xi _A$
describes the possible deviation of the cosmic ray drift velocity from the gas velocity.
Using Eq. (5) we obtain
\begin{equation}
w=u+\xi _A\frac {\dot{R}_f}{M^f_A}\sqrt{\frac {\rho }{3\rho _0}}, \ r>R_c
\end{equation}
The similar expression for the cosmic ray drift velocity is used upstream of
the reverse shock at $r<R_b$.
We  use values $\xi _A=1$ and $\xi _A=-1$ upstream of the
forward and reverse shocks respectively, where Alfv\'en waves are
generated by the cosmic ray streaming instability and propagate in
the corresponding directions. The damping of these waves heats the gas
upstream of the shocks (see McKenzie \& V\"olk \citealp{mckenzie82})
and limits the total compression ratios by a number close
to 6.
In the downstream region of
the forward and reverse shock at $R_b<r<R_f$ we put $\xi _A=0$ and therefore $w=u$.

Here we  use the energy dependence of the CR diffusion coefficient
like in the case of Bohm diffusion: $D=\eta _BD_B$ calculated for
the magnetic field  radial dependencies given by Eq. (5) and (6).
The parameter $\eta _B$ describes the possible deviations of
diffusion coefficient from the one achieved in the regime of Bohm
diffusion, $D_B=vpc/3qB$. Since the highest energy particles are
scattered by small-scale magnetic fields, their diffusion is
faster than the Bohm diffusion (Zirakashvili \& Ptuskin
\citealp{zirakashvili08}). The same is true for low  energy
particles because they can be resonantly scattered only by a
fraction of the magnetic spectrum.   Throughout the paper we use
the value $\eta _B=2$.

In real situations  the level of MHD turbulence drops with distance
upstream of the shock, so  the diffusion could be quite fast there. The
characteristic diffusive scale of highest energy particles is
 a small fraction $\xi _0<<1$ of the shock radius (see Zirakashvili \& Ptuskin
\citealp{zirakashvili08}) and is determined by the generation and
transport of MHD turbulence in the upstream region (see also
Vladimirov et al. \citealp{vladimirov06}, Amato \& Blasi
\citealp{amato06}). The value $\xi _0\sim \ln ^{-1}(D_c/D_s)$ is determined by ratio of diffusion
coefficient $D_c$ in the circumstellar medium and diffusion coefficient $D_s<<D_c$ in the vicinity of the
shock. The MHD turbulence is amplified exponentially in time
before the shock arrival from the background
level by cosmic ray streaming instability.
Since this process is not modelled here, we
simply multiply the CR diffusion coefficient $D$ to  the additional
factor $\exp ((\xi _1-1)/\xi _0)$ upstream of the forward shock and to  a  similar factor
$\exp ((1-\xi _4)/\xi _0)$ upstream  of the reverse shock. The characteristic range of $\xi _0$ is
$0.05\div 0.1$ (Zirakashvili \& Ptuskin \citealp{zirakashvili08}).

It is believed that the supernova ejecta has some velocity distribution $P(V)$ just after
the supernova explosion
(e.g. Chevalier \citealp{chevalier82b})
\begin{equation}
P(V)=\frac {3(k-3)}{4\pi k}
\left\{ \begin{array}{ll}
1, \ V<V_{ej} \\
\left( V/V_{ej}\right) ^{-k}, \ V>V_{ej}.
\end{array} \right.
\end{equation}
Here the index $k$ characterizes the steep power-low part of this distribution.
The radial distribution of
ejecta density is described by the same expression with $V=r/t$. The characteristic ejecta
velocity $V_{ej}$ can be expressed in terms of energy of supernova explosion $E_{SN}$ and ejecta mass $M_{ej}$
as

\begin{equation}
V_{ej}=\left( \frac {10(k-5)E_{SN}}{3(k-3)M_{ej}}\right) ^{1/2}
\end{equation}

Figures 1-5  illustrate the numerical results that are
obtained for the SNR shock propagating in the medium with hydrogen number density
 $n_H=0.09$ cm$^{-3}$ and temperature $T=10^4$ K.
The number density  of helium nuclei $0.1n_H$ was assumed.
We use the ejecta mass
 $M_{ej}=1.5M_{\odot }$,  the energy of explosion $E_{SN}=2.7\cdot 10^{51}$ erg and the parameter of ejecta
velocity distribution $k=7$.
These combination of these parameters is chosen to explain the obtained gamma-ray fluxes
within the hadronic  model of gamma-rays    (see next Section).

The initial forward shock velocity is
 $V_0=4.6\cdot 10^4$ km s$^{-1}$. The injection efficiency is taken to be
independent of  time $\eta _b=\eta _f=0.01$, and the injection
momenta are $p_{f}=2m(\dot{R}_f-u(R+0,t))$, $p_{b}=2m(u(R_b-0,t)-\dot{R}_b)$.
The  high injection efficiency
results in significant shock modification already at early epochs of the SNR expansion while the
thermal sub-shock compression ratio is close to 2.5.  This is in agreement with
calculations  of  collisionless shocks (Zirakashvili \citealp{zirakashvili07}) and are supported
by  radio-observations of young extragalactic SNRs
(Chevalier \citealp{chevalier06}).  In hadronic models of gamma-radiation of RX J1713.7-3946
the  ratio  of relativistic electrons to protons  $K_{ep}$ appears  significantly lower
compared to the observed ratio for galactic cosmic rays which is around  0.01 at 10 GeV. A possible
explanation could be the dependence of the electron injection efficiency on the shock speed.
In the paper we assume that   the electron injection is  inversely proportional to
the square of the shock speed. Then the
electron to proton ratio $K_{ep}\sim 10^{-4}$ for a characteristic shock speed  3000 km s$^{-1}$ of  young
SNRs corresponds to $K_{ep}\sim 10^{-2}$ in the old SNRs with a characteristic shock speed
300 km s$^{-1}$. The old SNRs probably produce the  main part of galactic GeV electrons.

\begin{figure}[t]
\includegraphics[width=8.0cm]{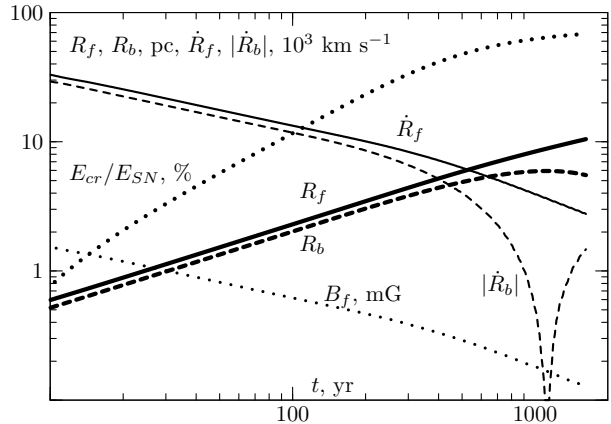}
\caption{Time dependencies of parameters characterizing the forward and reverses shocks:
the forward shock radius $R_f$ (thick solid line),
the reverse shock radius $R_b$ (thick dashed line), the forward shock velocity $\dot{R}_f$
 (thin solid line); the reverse shock velocity $\dot{R}_b$
 (thin dashed line); the magnetic field strength downstream of the forward shock (thin dotted line);
the ratio of the CR energy to the  total energy of the supernova
explosion  $E_{cr}/E_{SN}$ (dotted line).}
\end{figure}


The time evolution of the shock radii $R_f$ and $R_b$,  the forward and
reverse shock speeds  $V_f=\dot{R}_f$ and $V_b=\dot{R}_b$,  the CR energy
$E_{cr}/E_{SN}$ and the magnetic field strength $B_f$  downstream of the forward shock  are shown in Fig.1.
The calculations were performed until the present epoch at  $t=1620$ yr, when the
forward shock speed  drops down to $\dot{R}_f=2.76\cdot 10^3$ km s$^{-1}$
and the forward shock radius $R_f=10.5$ pc  corresponding to the
angular radius $30'$ of the remnant RX J1713.7-3946, assuming 1.2 kpc distance to the source.
At early epochs  of SNR evolution the distance between reverse and
forward shocks is only 10$\% $ of the remnant radius. Note that the
automodel Chevalier-Nadezhin solution with $k=7$   predicts  23$\% $ thickness (Chevalier \citealp{chevalier82}).
This can  be attributed
to the  strong modification of both shocks by CR pressure. The
reverse shock is strongly decelerated only when the forward shock
sweeps the gas mass comparable to the ejecta mass at $t>100$ yr
and when the transition to the Sedov phase begins. It is important
that the observable ratio $R_b/R_f\sim 0.5$ is achieved at the Sedov
phase when the forward shock speed  is close to 1/3  of
the characteristic ejecta speed  $V_{ej}$. Since the synchrotron X-ray emission of SNR
RX~J1713.7-3946  extends well beyond 1 keV, the  present forward
shock speed  cannot be significantly less than   $3\cdot 10^3$ km s$^{-1}$.
Therefore the characteristic velocity of the ejecta  cannot be smaller than
$10^4$ km s$^{-1}$. Such high velocities may be
attributed to Ib/c and IIb core collapse supernovae with low
ejecta masses, but not to the most frequent core collapse IIP
supernovae with characteristic ejecta velocities $3\cdot
10^3-4\cdot 10^3$ km s$^{-1}$ and high ejecta masses.

\begin{figure}[t]
\includegraphics[width=8.0cm]{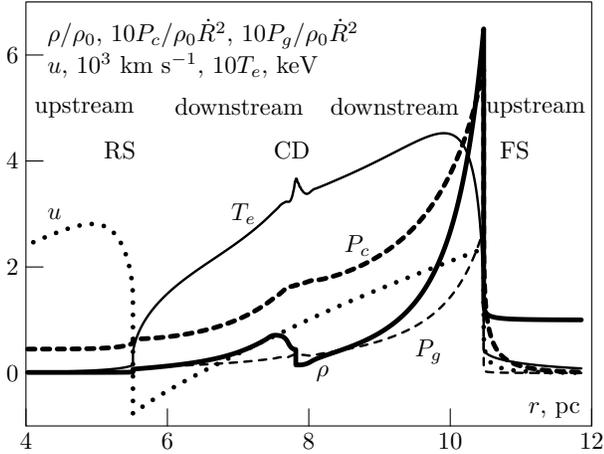}
\caption{Radial dependencies of the gas density (thick solid line), the gas
velocity (dotted line), CR pressure (thick dashed line), the gas pressure (dashed line) and
the electron temperature (thin solid line) at
 $t=1620$ yr.  The calculations result in the following parameters in the present epoch:
 the forward shock velocity 2760 km s$^{-1}$, its radius 10.5 pc,
the magnetic field strength downstream of the forward shock 127 $\mu $G.
In the same figure we show the positions of the forward and reverse  shocks,
(FS and RS, respectively) and the contact discontinuity (CD). }
\end{figure}

Radial dependencies of  several key parameters  at the present epoch
 $t=1620$ yr are shown in Fig.2.   In the same figure we show the positions  of
the contact discontinuity and the forward and reverse shocks.
The contact discontinuity between the ejecta and the interstellar gas
is located at $r=R_c=7.8$ pc. The reverse shock in the ejecta is situated at
$r=R_b=5.5$ pc. At the Sedov stage the reverse shock moves in the negative direction
and reaches  the center three thousand  years after the supernova explosion.
It should be noted that our one-dimensional calculations cannot
adequately describe the development of the Rayleigh-Taylor
instability of the contact discontinuity. In real situations
the supernova ejecta and the circumstellar gas are mixed by
turbulent motions in this region (see e.g. MHD modeling of Jun \&
Norman \citealp{jun96}).

\begin{figure}[t]
\includegraphics[width=8.0cm]{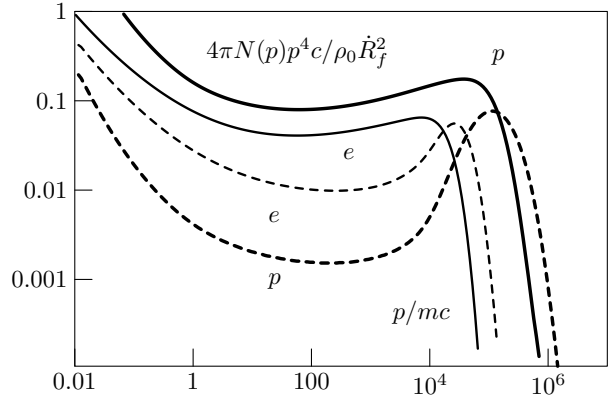}
\caption{The energy distributions of  accelerated protons (thick lines) and electrons
(multiplied to the factor of 5000, thin lines) at the epoch $t=1620$ yr.
The spectra at both the forward shock (solid lines) and at the reverse shock (dashed lines) are shown. }
\end{figure}

\begin{figure}[t]
\includegraphics[width=8.0cm]{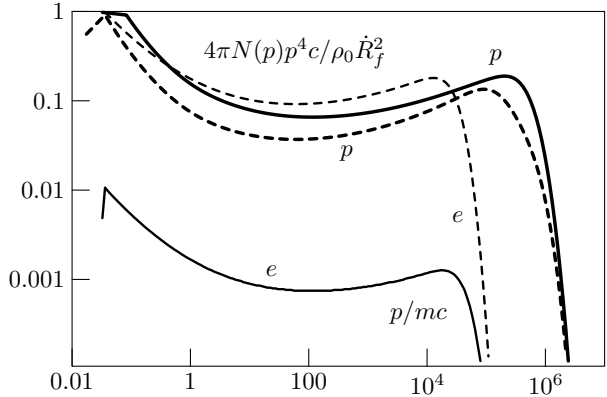}
\caption{The energy distributions of accelerated protons (thick lines) and
electrons (multiplied to the factor of 5000, thin lines) at $t=100$ yr.
Spectra at both the forward shock (solid lines) and the reverse shock (dashed lines) are shown.}
\end{figure}

In  SNRs the plasma is heated to keV temperatures.
The thermal X-ray emission is determined by the electron temperature.
In young  SNRs thermal electrons are not in equilibrium with protons if the
temperature equilibration proceeds through the  minimum available electron heating, i.e.
through  Coulomb collisions (Spitzer \citealp{spitzer68}). The evolution
of electron temperature $T_e$ is described by the following equation

\[
\frac {\partial T_e}{\partial t}=-u\frac {\partial T_e}{\partial r}
-\frac {2T_e}{3r^2}\frac {\partial r^2u}{\partial r}
\]
\begin{equation}
+P_g\frac {8\sqrt{2}q^4}{3m_em} \left( \frac {m_e}{T_e}\right) ^{3/2}\lambda _{ep}
\end{equation}
where $\lambda _{ep}\sim 30$ is the Coulomb logarithm. The corresponding temperature is
also shown in Fig.2.

\begin{figure}[t]
\includegraphics[width=8.0cm]{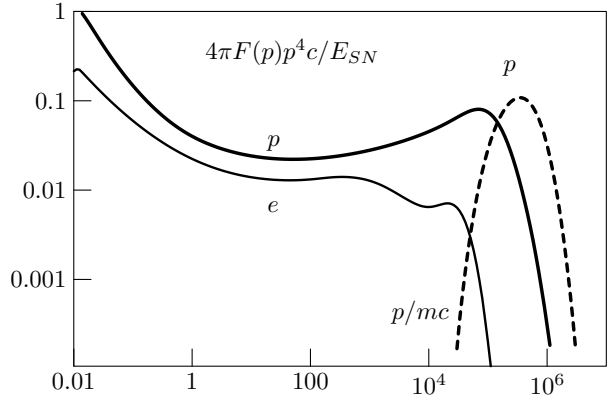}
\caption{Spatially integrated spectra of accelerated protons (solid line) and electrons
(multiplied to the factor of 5000, thin solid line) at $t=1620$ yr. Spectrum of run-away particles which have
left the remnant is also  shown (dashed line). }
\end{figure}

Spectra of accelerated protons and electrons are shown in Fig.3.
At the present epoch  the maximum energy of protons accelerated
in this SNR is about 140 TeV, while higher energy particles have
already left the remnant. The spectra of both electrons and protons at the reverse shock show
significant  bumps just before the cut-offs. Because  the reverse
shock presently moves in the rarefied medium of the ejecta,
the amount of freshly injected  low-energy particles is relatively
small, while many high energy particles accelerated earlier have not left
the reverse shock yet, and continue to be accelerated.

The spectra for  an early epoch, $t=100$ yr are shown in Fig.4. Since the
shock speed  was $V_f=1.3\cdot 10^4$ km s$^{-1}$, the maximum energy of particles was
also higher. For the forward shock it is about 650 TeV.

Spatially integrated proton and electron spectra at the present
epoch  $t=1620$ yr are shown in Fig.5. We also show the spectrum
of run-away particles. These particles have already left the
simulation domain through an absorbing boundary at $r=2R_f$. The
sum of the proton spectra shown  is the total  cosmic ray  spectrum 
produced in this SNR  over the last 1620 years after SN explosion.
For this SNR, the spectrum of  cosmic ray protons have  a maximum
at 800 TeV. This is somewhat smaller than the required value to
explain the knee in the cosmic ray spectrum.

The  results of calculations  of radiation produced by accelerated electrons and protons
are presented and discussed in  the next sections.
The gamma-ray spectra   from proton-proton interactions  are calculated using the formalism of
Kelner et al. \citealp{kelner06}.  For calculations of  IC gamma-rays we use  standard
expressions (see e.g. Blumenthal \& Gould \citealp{blumenthal70}). In calculations we
take into account  the target photons of the microwave background radiation only. We checked that
other diffuse radiation fields do not  contribute significantly to the gamma-ray production,
unless the SNR is located in a regions  with significantly (by an order of magnitude)
enhanced optical and infrared radiation background.
For calculations of synchrotron radiation we take into account that in real
situations the magnetic field has  some  probability distribution $P(B)$.
This makes the cut-off in  the spectrum of synchrotron radiation
somewhat smoother (see Appendix).

\begin{figure*}[t]
\includegraphics[width=14.0cm]{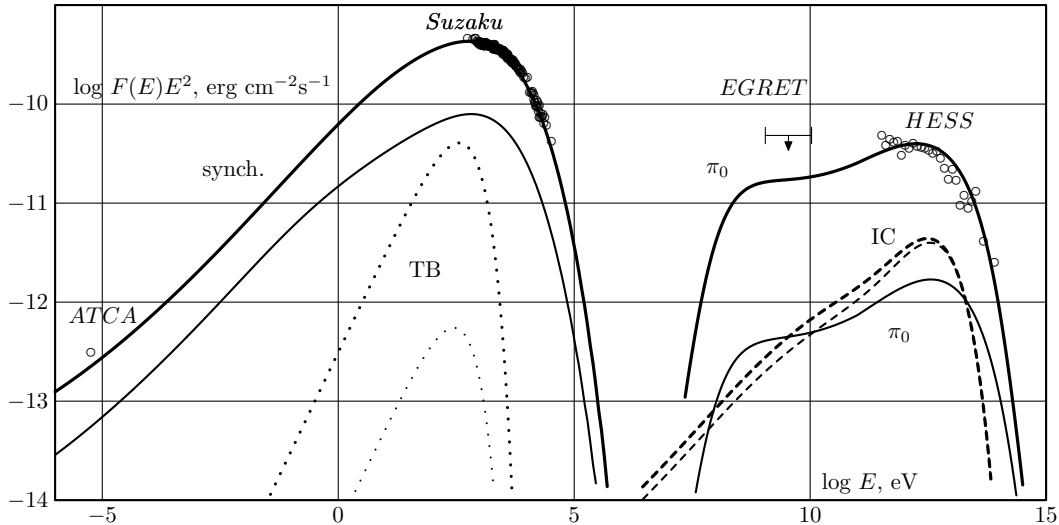}
\caption{The results of modeling of of nonthermal radiation of RX J1713.7-3946
within the hadronic scenario of gamma-ray production. The following basic
parameters are used: $t=1620$ yr, $D=1.2$ kpc, $n_H=$0.09 cm$^{-3}$,
$E_{SN}=2.7\cdot 10^{51}$ erg, $M_{ej}=1.5M_{\odot}$, $M^f_A=M^b_A=23$, $\xi _0=0.05$,
the electron to proton ratios at the forward and reverse shocks $K^f_{ep}=
10^{-4}$ and $K^b_{ep}=1.4\cdot 10^{-3}$.
The calculations lead to the following values of the magnetic fields and the shock speeds at
the present epoch: the magnetic field downstream of the forward and reverse shocks
$B_f=127$ $\mu $G and  $B_b=21$ $\mu $G respectively, the speed of the
forward shock $V_f=2760$ km s$^{-1}$,  the
speed of the reverse shock  $V_b=-1470$ km s$^{-1}$. The following
radiation processes are taken into account: synchrotron radiation
of accelerated electrons (solid curve on the left), IC emission
(dashed line), gamma-ray emission from pion decay (solid line on
the right), thermal bremsstrahlung (dotted line). The input of the reverse shock is
shown by the corresponding thin lines.
 Experimental
data in  gamma-ray  (HESS; Aharonian et al. \citealp{aharonian07}) and
X-ray bands (Suzaku; Tanaka et al. \citealp{tanaka08}), as well as the  radio flux
$22\pm 2$ Jy at 1.4GHz (ATCA; Acero al. \citealp{acero09}) from the whole remnant  are also shown. }
\end{figure*}

\begin{figure*}
\includegraphics[width=14.0cm]{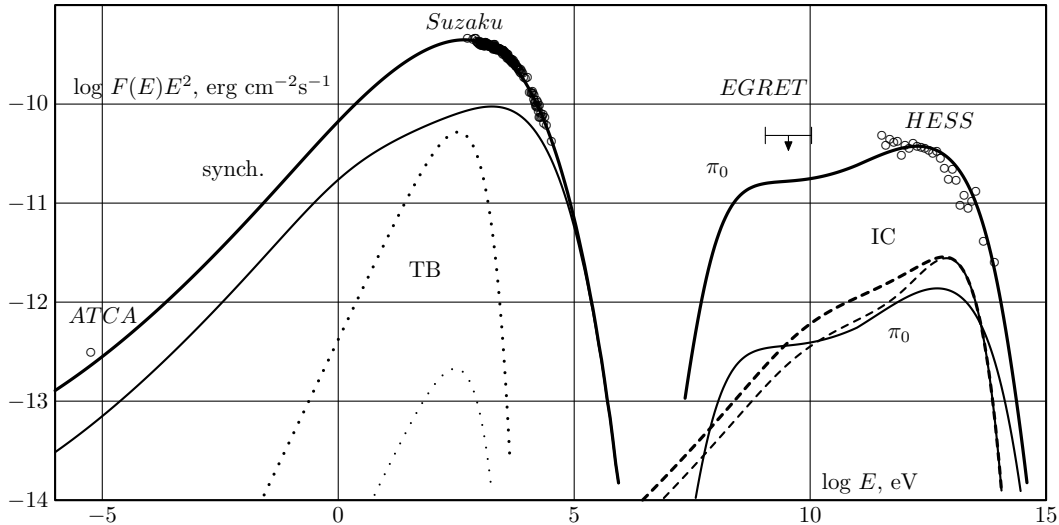}
\caption{The same as in Fig.6, but  for  the supernova shock
propagating in the medium with $r^2$ density profile, assuming the following parameters
 $t=1620$ yr, $D=1.3$ kpc,  $E_{SN}=2.5\cdot 10^{51}$ erg, $M_{ej}=1.4M_{\odot}$,
$M^f_A=M^b_A=23$, $\xi _0=0.05$, $K^f_{ep}=1.5\cdot 10^{-4}$, $K^b_{ep}=4.4\cdot 10^{-4}$.
The undisturbed hydrogen number density at the present
shock position is $n_H=$0.12cm$^{-3}$.
The calculations lead to the following values of the magnetic fields and the shock speeds at
the present epoch:
the magnetic field downstream of the forward and reverse shocks
$B_f=131$ $\mu $G and $B_b=28$ $\mu $G, respectively, the speed of the
forward shock $V_f=2260$ km s$^{-1}$, the
speed of the reverse shock $V_b=-3010$ km s$^{-1}$.  }
\end{figure*}

\section{Hadronic scenario}

The results of  calculations of the broad-band emission for the case of hadronic origin
of high energy gamma-rays  are shown  in Fig.6. The principal  model parameters used in
calculations are described in the figure caption.  Note that at the present epoch already
70 \% of the explosion energy of $2.7 \times 10^{51}$~erg
has been  transferred to accelerated protons (see Fig.1), most of them still confined
in the shell of the remnant
(see Fig.5).   The maximum energies are  140~TeV and 23~TeV for  protons and electrons respectively.
Although the reverse shock contributes significantly to the highest energy particles,
especially around 100 TeV (see Fig.3), the gamma-ray production related to this population of
protons is suppressed
because of the low density of the ejecta plasma. The contribution of electrons directly accelerated by the
reverse shock to synchrotron radiation and IC gamma-rays is more significant.  In particular,
the reverse shock produces 16$\% $ of X-rays.  Moreover, the
bump at the end of IC spectrum is contributed mostly  by these  electrons.
The injection efficiency of electrons is adjusted in order to reproduce the total
intensity of synchrotron X-rays.  The electron injection efficiency at the reverse
shock was adjusted  to reproduce the observable radio-intensity of the inner ring. The gray scale radio
images of Ellison et al. \citealp{ellison01} and Lazendic et al. \citealp{lazendic04} were used in order to obtain
a rough estimate of radio-flux 4 Jy at 1.4 GHz from this region.  The modelled total
radio flux is slightly below the observed flux. Note however that the space-integrated
faint radio flux  contains uncertainties, in particular the contribution of the background
thermal radio emission can be quite  significant.

In Fig.6 we show also the energy flux of the  thermal bremsstrahlung. It has a
maximum at 0.4 keV and approximately equals to the energy flux of
gamma-rays  produced  in proton-proton interactions .
Note that the ratio of the energy flux of the thermal
emission to the flux of  "hadronic" gamma-rays does not depend on the density of
ambient plasma, but  depends on the
shock  speed  and the cosmic ray acceleration efficiency (see  Appendix).

It is expected that a circumstellar medium around progenitors of
Ib/c and IIb supernova is strongly nonuniform. At the main
sequence (MS) phase the stellar wind of progenitor creates
rarefied bubble in the surrounding medium. Later the part of this
bubble is filled with a dense gas ejected by progenitor at the Red
Super Giant (RSG) phase of the stellar evolution. At the end of the
progenitor evolution this gas most likely will be  swept up and transported
to the MS bubble periphery by a powerful Wolf-Rayet stellar wind.
(see Chevalier \citealp{chevalier05} for a review).

\begin{figure*}[t]
\includegraphics[width=14.0cm]{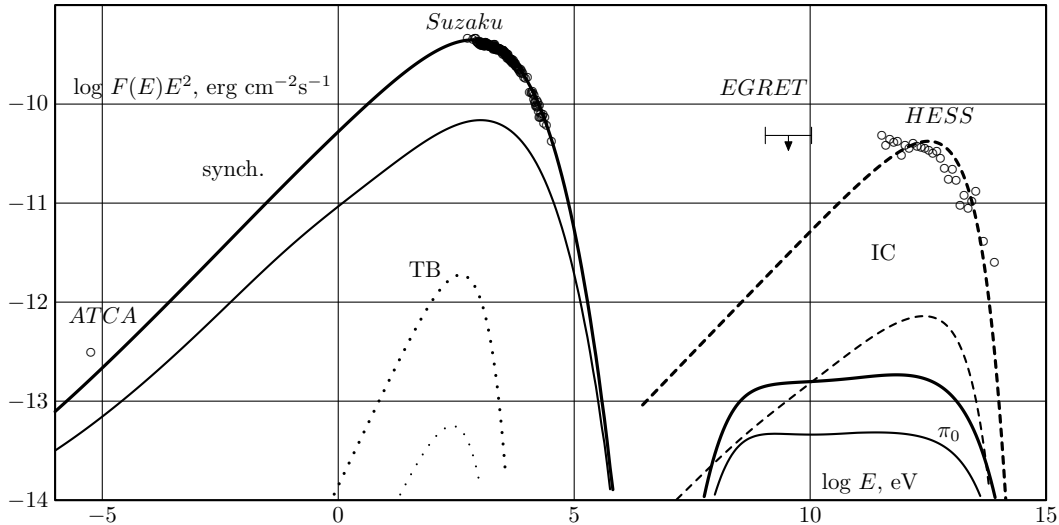}
\caption{Broad-band  emission of RX J1713.7-3946  for the
leptonic scenario of  gamma-rays with
a non-modified forward shock.  The principal model parameters are:
$t=1620$ yr, $D=1.5$ kpc, $n_H=$0.02 cm$^{-3}$, $E_{SN}=1.2\cdot 10^{51}$ erg,
$M_{ej}=0.74M_{\odot}$, $M^f_A=69$, $M^b_A=10$, $\xi _0=0.1$,
$K^f_{ep}=2.3\cdot 10^{-2}$, $K^b_{ep}=9\cdot 10^{-4}$.
The calculations lead to the following values of the magnetic fields and the shock speeds at
the present epoch:
the magnetic field downstream of the forward and reverse shocks $B_f=17$ $\mu $G and $B_b=31$ $\mu $G,
respectively, the speed of the
forward shock
$V_f=3830$ km s$^{-1}$, the
speed of the reverse shock $V_b=-1220$ km s$^{-1}$. The following
radiation processes are taken into account: synchrotron radiation
of accelerated electrons (solid curve on the left), IC emission
(dashed line), gamma-ray emission from pion decay (solid line on
the right), thermal bremsstrahlung (dotted line). The input of the reverse shock is
shown by the corresponding thin lines.}
\end{figure*}

\begin{figure*}
\includegraphics[width=14.0cm]{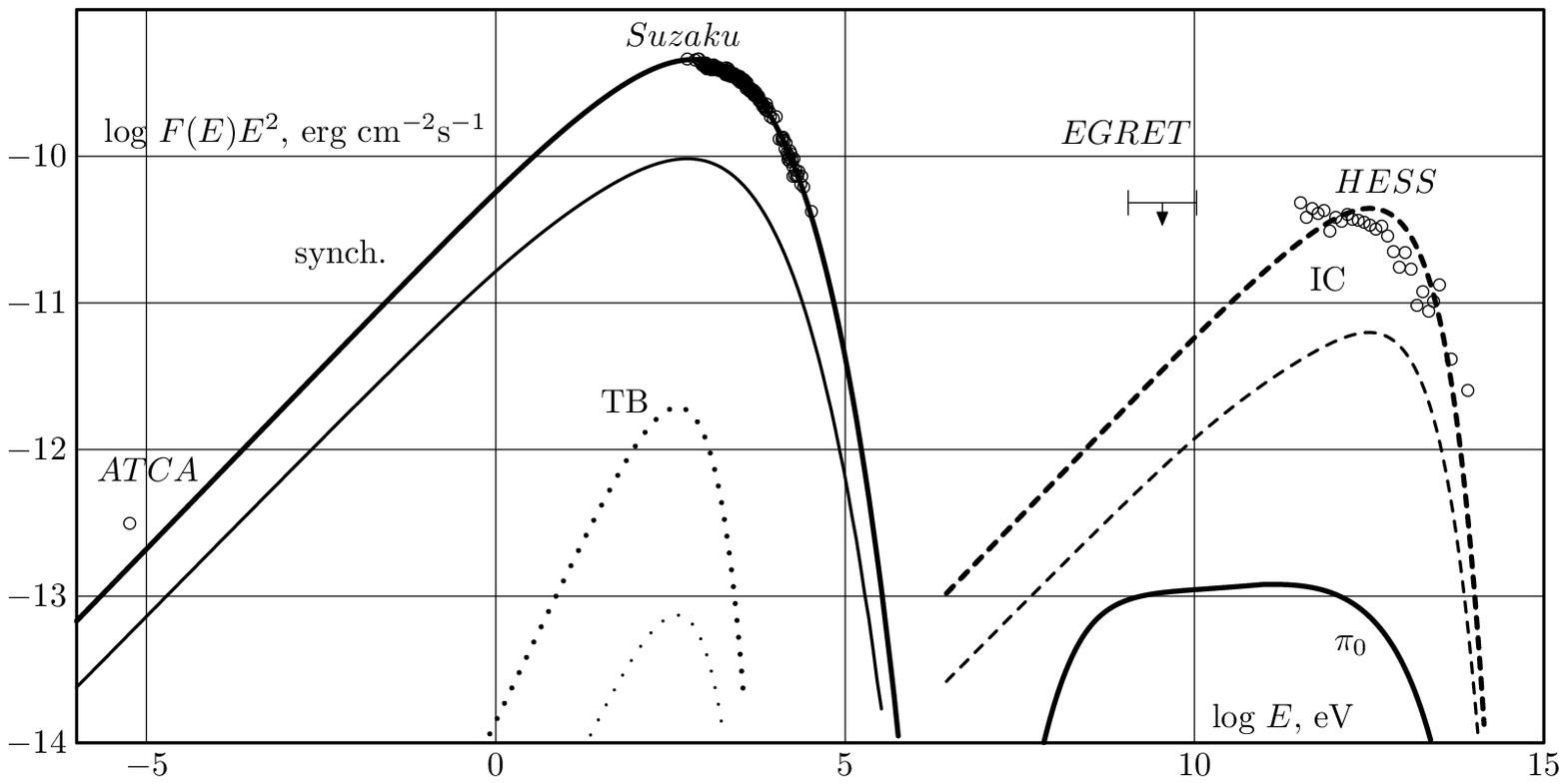}
\caption{The same as in Fig.8  but for
unmodified forward and reverse shocks.  The principal model parameters are:
$t=1620$ yr, $D=1.5$ kpc,
$n_H=$0.02 cm$^{-3}$, $E_{SN}=1.25\cdot 10^{51}$ erg,
$M_{ej}=0.82M_{\odot}$, $M^b_A=23$, $M^f_A=69$, $\xi _0=0.1$,
 $K^f_{ep}=2.0\cdot 10^{-2}$, $K^b_{ep}=6.8\cdot 10^{-2}$.
The calculations lead to the following values of the magnetic fields and the shock speeds at
the present epoch:
the magnetic field downstream of the forward and reverse shocks
$B_f=17.5$ $\mu $G and $B_b=13$ $\mu $G, respectively, the speed of the
forward shock
$V_f=3870$ km s$^{-1}$, the
speed of the reverse shock $V_b=-1290$ km s$^{-1}$. }
\end{figure*}

It seems that the corresponding MS bubble is indeed found.
SNR RX J1713.7-3946 is surrounded by a massive shell of
molecular gas (Fukui \citealp{fukui08}). The densest cores of
molecular gas (clouds C and D) are probably swept up by the
forward shock of the SNR. This is the reason to consider the SNR
evolution in the circumstellar medium with a positive density gradient. The results obtained for
the gas density profile proportional to the square of the radius are shown in Fig.7.
A similar "bubble" model, but  with a sharper density profile
was considered by Berezhko \& V\"olk \citealp{berezhko06}.

Results of  calculations of the broad-band emission presented in Fig.7 are quite
similar to the ones shown in Fig.6, however the agreement with observations is achieved with
different set of model parameters. Also, note that in this case the
reverse shock produces  35$\% $ of the total X-ray emission.

Note that in both cases the results are obtained for a low density gas.   A
similar number density close to $n_H\sim 0.1$ cm$^{-3}$  has been suggested
by Morlino et al. \citealp{morlino08} in their hadronic models of gamma-emission
of SNR RX J1713.7-3946.  The low density gas  is required also  from the
upper limit on the thermal X-ray emission from this supernova remnant (see e.g.
Cassam-Chena\"i et al. \citealp{cassam04},  Takahashi et
al. \citealp{takahashi08}, Tanaka et al. \citealp{tanaka08})  and is in
agreement with a simple estimate of the gamma-ray intensity
(see Appendix C).  Significantly denser gas,
$n_H\sim 1$ cm$^{-3}$,  was assumed  by Berezhko \& V\"olk \citealp{berezhko06}. This is because
they assumed that the cosmic ray acceleration occurs only at 1/5 part of the shock surface.

\section{Leptonic scenario}

The observed X-ray to gamma-ray  energy flux ratio close to 15 determines the  average
value of the magnetic field  $B\sim 12$ $\mu $G when the leptonic origin of gamma-emission
in RX J1713.7-3946 becomes possible. How realistic is such a week magnetic field?

The  amplified magnetic fields  can be quite  weak if the SNR shock propagates in the
rarefied medium.  However since the magnetic field at the reverse shock is several times
lower than the magnetic field at the forward shock (see previous Section) and the X-ray
intensities of the shocks are comparable, the IC emission will be mainly produced at the
reverse shock; this seems to be in conflict  with HESS observations.

For this reason we consider a unmodified forward shock of SNR.
If the injection is not effective at the forward shock, the energy density of
cosmic rays will be low and the same will be true for
the amplified magnetic field. Such a situation is possible, in particular,  for a perpendicular
SNR shock propagating in the medium with an azimuthal magnetic field.
In this case we use a very  small injection parameter $\eta _f=10^{-5}$ at the
forward shock and the standard value $\eta _b=10^{-2}$ for the
injection parameter at the reverse shock. Thus in this scenario we deal with
unmodified forward shock and modified reverse shock.
Since, under these assumption,  the electron to proton ratio
$K_{ep}$ is close to the  observed ratio for galactic cosmic rays,  0.01 at 10 GeV,
the electron injection was taken to be independent on the shock velocity.
We also use a higher value $M^f_A=69$ for the forward shock, and adopt the
distance $D=1.5$ kpc that is close to an upper limit according to
Cassam-Chena\"i et al. \citealp{cassam04} and Fukui \citealp{fukui08}.

We should note that the shock speed is one of the key parameters in order to achieve  the good feet.
The shock speed is related with a remnant size at given age of the remnant. This explains a slight
difference of the assumed value of distances.

The results are shown in Fig.8. The magnetic fields  at the reverse shock
and the forward shock are comparable in this case. The energy of cosmic rays
is only 10$\% $ of the explosion energy. The cosmic ray pressure at the forward shock
is only 2$\% $ of the ram pressure $\rho _0\dot{R}^2_f$. The particles are mainly accelerated
at the reverse shock.  The maximum energies are 55 TeV and 36 TeV for protons and electrons respectively.
The reverse shock produces 16 $\% $ of X-ray emission.

For completeness we  also consider the case when both forward and
reverse shocks are not modified. The injection efficiency $\eta
_b=\eta _f=10^{-5}$ is used. The results are shown in Fig. 9. The
energy of cosmic rays is 3$\% $ of the explosion energy. The
reverse shock produces 20$\% $ of X-ray emission.

\section{Radial profiles}

\begin{figure}
\includegraphics[width=8.0cm]{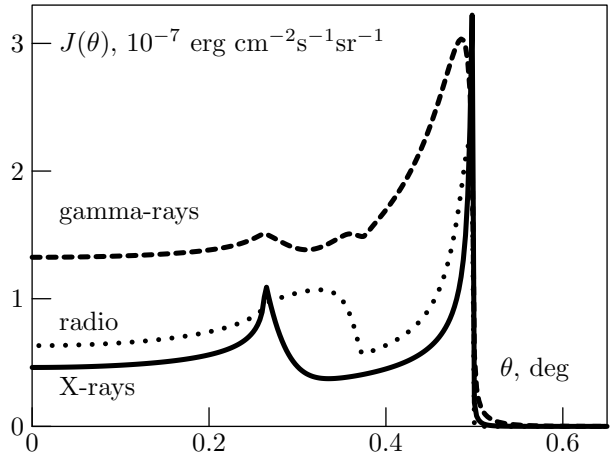}
\caption{Radial profiles of 2 keV X-rays  (multiplied to 0.04, solid line),
1 TeV gamma-ray emission (dashed line)
and 1.4 GHz radio-emission (multiplied to  $10^3$, dotted line),  calculated for the hadronic
scenario in the uniform medium.}
\end{figure}

The radial profiles of brightness distributions of  X-ray, gamma-ray  and radio- emissions
calculated  for leptonic and hadronic models are shown
in Fig.10-12. The projection effect is  taken into account.

\begin{figure}
\includegraphics[width=8.0cm]{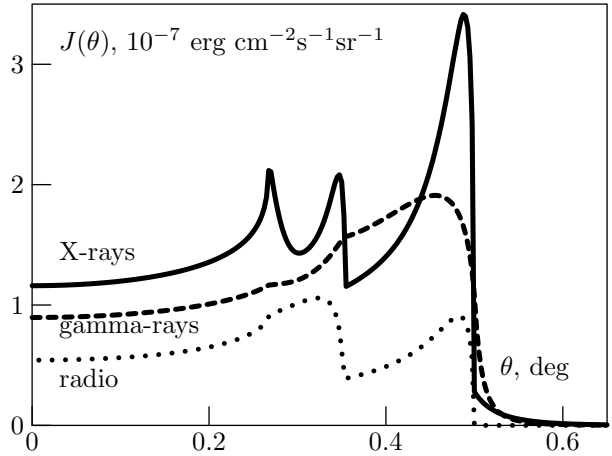}
\caption{Profiles of 2 keV X-ray emission (multiplied to 0.1, solid line),
1 TeV gamma-emission (dashed line)
and 1.4 GHz radio-emission (multiplied to  $10^3$, dotted line) for the leptonic
scenario with the non-modified forward shock.}
\end{figure}

\begin{figure}
\includegraphics[width=8.0cm]{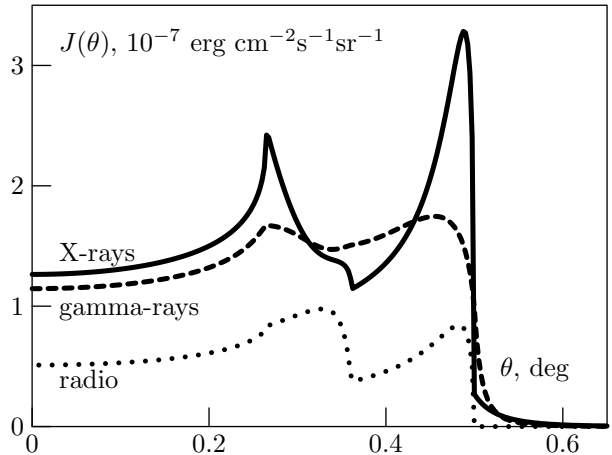}
\caption{Radial profiles of 2 keV X-rays (multiplied to 0.1, solid line),
1 TeV gamma-ray emission (dashed line)
and 1.4 GHz radio-emission (multiplied to  $10^3$, dotted line) for the leptonic
scenario with the non-modified forward and reverse shocks.}
\end{figure}

\begin{figure}[t]
\includegraphics[width=8.0cm]{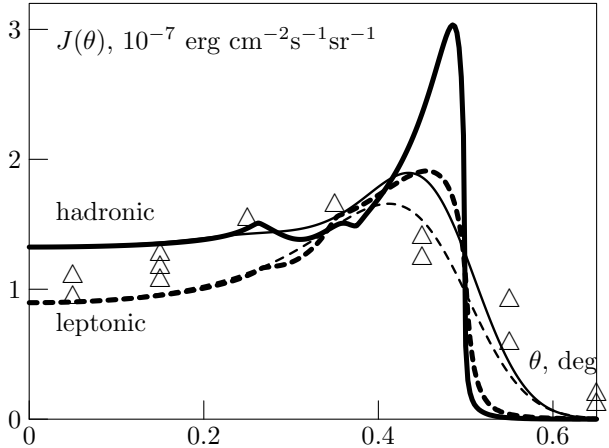}
\caption{Radial Profiles of 1 TeV gamma-rays  for the hadronic scenario in the uniform medium
 (solid) and for the leptonic scenario with the unmodified forward shock (dashed).
The profiles smoothed with a Gaussian point spread function with
$\sigma =0.05^{\circ }$ are  also shown (thin lines). The triangles show the  azimuthally averaged
TeV gamma-ray  radial profile observed by HESS
(Aharonian et al. \citealp{aharonian07}).}
\end{figure}

For the hadronic  scenario in the  uniform medium,  all
three components of radiation in the radio, X-ray and gamma-ray
bands  peak at the forward shock (approximately $0.5^\circ$ of the
angular radius.  Note that the synchrotron emission has a sharper
feature at the forward shock than the gamma-ray emission. The
synchrotron emission also shows a noticeable peak at the reverse
shock  (approximately 1/3 of the brightness at the forward shock).
The slight feature of gamma-rays  (at the level of 10 \%) at the
reverse shock is due to the IC gamma-rays. The middle peak of
X-ray emission in Fig. 11 is explained by the electrons
accelerated at the forward shock and reaching the contact
discontinuity with a stronger magnetic field. In Fig. 13 the
calculated gamma-ray profiles are compared with the  HESS data.

The most prominent feature at all X -ray radial profiles from Figs.~10-12
is a clear visible filament at the reverse shock.  Note that the
projection effect works differently for reverse shock in
comparison with the forward shock (see Appendix D). As a result, at the reverse shock,
the filaments are visible even in the relatively weak magnetic
fields (13-30 $\mu G$). Such filaments have been
indeed observed by Chandra  in the inner rim  (Uchiyama et al.
\citealp{uchiyama03}, Lazendic et al. \citealp{lazendic04}).

The scale $l_d$ of an exponential decrease of the synchrotron emissivity downstream of the shock with
a velocity $V_s$ is
given by (Zirakashvili \& Aharonian \citealp{zirakashvili07})

\[
l_d=\frac {1.4\cdot 10^{17}\ \mathrm{\ cm}}{\sqrt{1+\kappa
^{1/2}}} \eta _B^{1/4}\left( \frac {V_s}{\mathrm{3\cdot 10^3\ km\ }s^{-1}}
\right) ^{1/2}\times
\]
\begin{equation}
\left( \frac {B_d}{100\ \mu \mathrm{G}}\right) ^{-3/2} \left( \frac
{\hbar \ \omega }{\mathrm{1\ keV\ }} \right) ^{-1/4} \ . 
\end{equation}

Here $\kappa =B_u/B_d$ is the ratio of magnetic fields upstream and downstream of the shock.
The only available profile of the linear filament in the inner region is given by Lazendic et al.
\citealp{lazendic04}.
Note that the profile is probably slightly influenced by CCD gap of Chandra.
The filament
width at the half of magnitude is about 40$''$. This corresponds to 30$''$ for the exponential scale
(see Appendix D) or $l_d=7\cdot 10^{17}$ cm at $D=1.5$ kpc. Using the last formula we find $B_d=25\ \mu G$ .
So the
magnetic field at the reverse shock in our calculations  corresponds to the width of X-ray filaments.
According to Fig.10,  filaments with a similar width must be also observed
at the outer X-rims in the hadronic model.

\section{Maximum energies and magnetic amplification at the reverse and forward shocks of SNR}

According to Zirakashvili \& Ptuskin \cite{zirakashvili08} the maximum energy of protons $E_m$ accelerated at
the quasi-parallel shock moving in the circumstellar medium with the magnetic field strength
$B_0$  and Alfv\'en velocity $V_a=B_0/\sqrt{4\pi \rho _0}$ may be found from the expression

\begin{equation}
E_m=21\ \mbox{TeV}\ \frac {\left( \frac {\eta _{esc}}{0.05}\right)
\left( \frac {V_f}{10^3\
\mbox{km\ s}^{-1}}\right) ^2
n_H^{1/2}m_{\exp }R^f_{pc} } { \ln \left( \frac
{2B_0}{B_b}\right) -1 +\left( 2\frac {u'^4}{u_*^4}-1\right)
^{1/4} }.
\end{equation}
Here $\eta _{esc}=2F_E/\rho _0V_f^3$ is the ratio of the energy
flux of run-away particles $F_E$ to the flux of kinetic energy
$\rho _0V_f^3/2$, $B_b$ is the strength of the random magnetic field
in the circumstellar medium, $u_*=1.73\cdot 10^3$ km s$^{-1}$. The
expansion parameter $m_{\exp }$ is determined as $m_{\exp }=d\ln
R_f/d\ln t$ and is close to 0.5 in supernova remnants in the
transition to the Sedov phase in the homogeneous medium. The speed
$u'$ is  a so-called normalized shock velocity

\begin{equation}
u'=V_f\ \left( \frac {V_a}{\mbox{10 km s}^{-1}}\right) ^{-3/4}
\left( \frac {\eta _{esc}}{0.05}\right) ^{1/2}
\end{equation}

Eq. (12) is  derived for the diffusive shock acceleration in the
presence of the non-resonant streaming instability suggested by
Bell \cite{bell04}. This equation is valid for $u'>u_*$. The maximum
energy for lower velocities when the resonant instability should be taken
into account is given by the same expression with the logarithm in the denominator only.
The ratio $\eta_{esc}=0.14$ for the  cosmic ray modified shock with the compression
ratio $\sigma =6$ and the compression ratio of the thermal
sub-shock 2.5. For magnetic field strength $B_0=5\mu $G in the
circumstellar medium, $V_f=2760$ km s$^{-1}$ from the hadronic
model (see Fig.6) and for a characteristic value of $\ln
2B_0/B_b=5$ we find $u'=1930$ km s$^{-1}$ and $E_m=136$ TeV. This
normalized shock velocity also determines the strength of the
amplified magnetic field $B\sim 2.5B_0=12.5\mu $G upstream of the
forward shock or $B\sim 75\mu $G downstream of the shock after the
compression in the shock transition region.

These numbers are in a qualitative agreement with our numerical
results obtained for the hadronic scenario. We found using Eq. (12)
that the maximum energies
are significantly (a factor of 40) lower in the leptonic scenario
with the non-modified forward shock. This is because a small
amount of accelerated protons. However the maximum  energy is
 underestimated if one uses Eq. (12) for an
oblique forward shock. The particles that run-away from such a
shock will produce a stronger electric current in a comparison
with the quasi-parallel case. Since the maximum energy $E_m$ is
proportional to the current which drives the streaming
instability this will permit to reach the maximum energy of
several tens TeV for the  leptonic scenario. This makes the
leptonic scenario with the non-modified forward shock to be
self-consistent.

The maximum energy and the strength of the amplified magnetic
field at the reverse shock depend on the magnetic field of
ejecta. The regular magnetic field strength $B\sim 100$ G of the
compact progenitor of Ib/c supernova with a radius $R=10^{12}$ cm
will drop down to $10^{-12}$ G after the homogeneous radial
expansion of the ejecta up to radius $3$ pc$\sim 10^{19}$ cm. Probably
inside the progenitor there exist stronger random magnetic fields
of the order $B\sim 10^4$ G. This fields will drop down to
$10^{-10}$ G after the homogeneous expansion. In the real
situation the expansion is not homogeneous and the random magnetic
field is stretched in the radial direction. This may additionally
amplify the field by a factor of $\sim 10$ that is the ratio of
the stellar radius to the expected azimuthal inhomogeneity scale
of the velocity field. This gives an upper limit $10^{-9}$ G for
the magnetic field strength of ejecta. The radial orientation of
the field is favorable for the ion injection at the reverse
shock.

The close value one may find in the model of
magneto-rotational supernova explosion of Bisnovatyi-Kogan et al.
\citealp{bisnovatyi-kogan08}. The magnetic field is amplified in the
differentially rotating iron core up to $B\sim 10^{17}$ G after
the core collapse and pushes the core material outward producing
the supernova explosion. For the core radius $10^{6}$ cm we found
$B\sim 10^{-9}$ G after the homogeneous expansion and $B\sim
10^{-8}$ G after inhomogeneous expansion. We shall use this upper
optimistic limit  $B_0\sim 10^{-8}$G for the estimate below.

In the hadronic model the reverse shock moves with the speed
$u=-4500$ km s$^{-1}$ in the ejecta frame of reference. The ejecta
number density $n^{ej}_H=9\cdot 10^{-4}$ cm$^{-3}$. Then the
Alfv\'en velocity $V_a=0.67$ km s$^{-1}$ and the normalized shock
velocity $u'=5.7\cdot 10^{9}$ km s$^{-1}$ for $\eta _{esc}=0.14$.
Then we find $E_m=2.3$ TeV using Eq. (12).  The magnetic
amplification factor was not determined by Zirakashvili \& Ptuskin
\citealp{zirakashvili08} for such high normalized velocities. The
amplification factor is 190 for the normalized velocity
$u'=3.9\cdot 10^{9}$ km s$^{-1}$. However we find $B\sim
300B_0=3\mu $G after an extrapolation. So the amplified field
downstream of the shock is $B\sim 18\mu $G. The maximum energy and
amplified magnetic field may be even higher if one takes into
account that the number density of the highest energy particles is
higher in comparison with what is expected at the shock with
run-away particles. These particles cannot run-away from the
reverse shock easily because the upstream region of the reverse
shock is located inside the remnant (see also bumps at the end of
the spectra in Fig.3). Amplified magnetic field weakly depends on
the initial seed field $B_0$ as $B\propto B_0^{1/4}$ while the
dependence of the maximum energy is stronger $E_m\propto
B_0^{3/4}$ for small $B_0$ (see Zirakashvili \& Ptuskin
\citealp{zirakashvili08} for details).

We conclude that for the magnetic field strength $B_0\sim 10^{-8}$G of ejecta the reverse shock may
accelerate particles up to multi-TeV energies and may amplify magnetic fields up to tens of
$\mu $G in the remnant considered.

\section{Discussion}

\subsection{Thermal X-ray emission}

The nonthermal emission of SNRs  cannot be treated out of the context of thermal emission. In particular,
any standard hadronic model of gamma-rays predicts  heating of the gas to high temperatures with
intense X-ray emission. The lack of such emission from the most pronounced TeV emitting SNR, RX J1713.7-3946,
can be interpreted as an argument against the hadronic models of gamma-rays
(Katz \& Waxman \citealp{katz08})  or an evidence of extremely high efficiency of transformation of the
kinetic energy of explosion to  acceleration of relativistic particles (Drury et al. \citealp{drury09}).

Though our numerical calculation show that  in the hadronic models of gamma-rays
the thermal bremsstrahlung appears to be  an order of magnitude below the Suzaku X-ray
data (see Figs 6 and 7), the actual flux of X-ray emission in this energy range is  higher
due to the contribution from the  line X-ray emission (see e.g. Ellison et al. \citealp{ellison07}).
The latter strongly depends on the chemical composition of the
X-ray emitting plasma. For the  composition similar to one observed in the Sun,
the intensity of  X-ray lines can significantly exceed the
free-free continuum  (see Ellison et al. \citealp{ellison07}). This means that
such lines must be observed contrary to X-ray data. Based on this fact,
Cassam-Chena\"i et al. \cite{cassam04} derived  an upper
limit on the ambient plasma density $n_H\sim 0.02$ cm$^{-3}$. This value is several times
lower than $n_H\sim 0.1$
 cm$^{-3}$ in the hadronic models considered in this paper.
 A simplest explanation of this could be  that  chemical composition of
the circumstellar gas is strongly different from the solar one e.g.
due to absorption of X-ray line emitting ions by dust grains.

We should note that the thermal X-ray emission may be strongly
depressed at the supernova shocks that are very strongly modified
by the cosmic ray pressure when the shock compression ratio
$\sigma >>10$. In this case only a small part of dissipating
energy is transferred to the thermal plasma and the gas
temperature may be below 0.1 keV even under the electron-ion
thermal equilibrium (Drury et al. \citealp{drury09}) However such
compression ratios are not  observed in our model. The strong
heating of the upstream plasma due to the damping of Alfv\'en
waves results in the modest shock modification with $\sigma \sim
6-7$. Even if one assumes that the waves are not damped  and  the
gas heating is negligible,  the wave pressure will be high and it
will also prevent the strong shock  modification (see Caprioli et
al. \citealp{caprioli09} for details). Very high compression
ratios are possible at radiative shocks in a high density plasma
where the waves are damped and the gas loses energy radiatively.

Another possibility to explain a low thermal X-ray luminosity of the remnant is related with a nonuniform
circumstellar medium. Chandra observations reveal a complex network of non-steady bright spots and
filaments inside the X-ray emitting shell (Uchiyama et al. \citealp{uchiyama07}). If dense filaments with
a small volume factor exist already in the circumstellar medium and determine the mean density of the
plasma,  say 1 cm$^{-3}$,  they will not be shocked by the SNR shock coming and
the mean density of the shocked material will be rather low say 0.01 cm$^{-3}$. However the plasma density
"seen" by TeV accelerated protons can be close to the total mean value because of the relatively fast
diffusion. As a result they will be able to produce  TeV gamma-ray with a flux close to the observed one.
An enhanced magnetic field inside the filaments may also produce the observed X-ray variability
of  bright  spots (Uchiyama et al. \citealp{uchiyama07}).

An interesting consequence  of this  scenario is that the gamma-ray intensity at low energies  can
be reduced  because of the slow diffusion of  GeV protons which prevents  their effective
penetration into the dense filaments. As a result, the gamma-ray spectrum due to p-p interactions
can be quite  similar to the IC  gamma-ray spectra expected in leptonic models.
This possibility makes quite difficult  the identification of the origin of  gamma-radiation.

It is interesting to note that the thermal X-rays produced in the gas shocked by the  forward shock are rare
observed in young supernova remnants. For example,  the thermal X-ray
emission of the supernova remnants Tycho and SN1006 is produced
by shocked ejecta that is enriched by heavy elements (Cassam-Chena\"i et al. \citealp{cassam07},
\citealp{cassam08}). In this regard the SNR RX J1713.7-3946 perhaps is not a very special example.
It is at a later evolutionary phase in comparison with SNRs mentioned above, so the ejecta is situated deeply
inside and has a very low density (see Fig.2).
This could be the  reason why the thermal X-ray emission of ejecta is not
observed. On the other hand,  the thermal emission from the downstream
region of the forward shock is not detected similar to many other young SNRs.

The young remnant RCW 86 is well known source  with thermal X-ray line emission observed
from the forward shock (see Yamaguchi  et al. \citealp{yamaguchi08}).  It is interesting to note that in this
remnant the chemical abundance of heavy ions downstream of the forward shock seems indeed below the solar one.

\subsection{Energy spectra and morphology}

The comparison of Figs. 6\&7 with Figs. 8\&9
shows that  the spectral shape of gamma-emission is better
reproduced in hadronic models. However the difference perhaps is
not sufficient for a  definite  conclusion  concerning the origin of
gamma-radiation radiation based on the current data.  The spectra predicted by
two models deviate significantly at  very high   ($\gg 10$~TeV) and low $\leq 10$~GeV
energies.  The first energy domain can be explored with the next generation
ground-based detectors with significantly improved sensitivity compared to HESS.
In particular,  the confirmation of the the last spectral points above 30 TeV reported by HESS
with limited statistical significance, would be a strong argument in favor of the hadronic
origin of gamma-rays.  Also,  the hadronic model predicts  gamma-ray flux
below 100 GeV at the level of  $\geq 10^{-11} \ \rm erg/cm^2 s$, while the  IC  gamma-ray
flux  predicted by the leptonic model  around 1 GeV  is closer to $10^{-12} \ \rm erg/cm^2 s$.
It is expected that the measurements conducted with the AGILE and Fermi telescopes
will soon  provide a decisive  answer  for the MeV/GeV  flux from RXJ~1713.7-3946.
The detection of a signal close to $10^{-11} \ \rm erg/cm^2 s$ would  imply hadronic origin of radiation.
On the other hand,   the  much lower flux would  favor the IC  origin of radiation but still could not
reject the hadronic origin, given that  gamma-rays could be produced in dense filaments.
As discussed above, because of slow diffusion of low energy protons that prevents their effective
penetration into the filaments,  the GeV emission  from p-p interactions in the dense filaments
could be suppressed.  This model can also give a reasonable, in our view,  explanation
of the lack of  thermal X-ray line emission.

We should note, however,  that the real picture could be more complex, in particular we
may expect non-negligible deviations of energy spectra because of  non-spherical
geometry, different shock properties at different parts of the shock surface,  {\it etc}.  Moreover,
one may expect  comparable  contributions  of gamma-rays from  interactions
of  both  accelerated protons and electrons, but produced in different proportions
in  different regions of the remnant  (see below).  In this regard,  detailed
spectroscopic measurements of gamma-rays from different parts of the remnant  seem to be a
key towards understanding of acceleration processes which take place in this source.

If the magnetic field is indeed as high as required in hadronic models,
we should expect, according to Fig.10,  thin filaments
with a width  of about  $30''$ in the remnant periphery.
The linear filaments are indeed present  at Chandra image
(Uchiyama et al. \citealp{uchiyama03}, Lazendic et al. \citealp{lazendic04})
but in the vicinity of the inner ring that might be related to  the reverse shock.
This agrees with Fig.10-12. The most recent X-ray image of XMM-Newton
(Acero et al. \citealp{acero09})  shows several linear filaments with the width
$1-2'$ present in  this image,  but they all are situated inside
the boundary of the  X-ray emission. The boundary itself is not sharp that is presumably related with a
nonuniform circumstellar medium.
The lack of sharp X-ray boundary is better explained  by leptonic models.
According to Fig.11 and Fig.12 some amount of  X-rays is  expected
from the region located beyond the forward shock. In the leptonic model it is possible
because accelerated electrons have sufficiently high energies to produce synchrotron
X-rays in the low magnetic field.

The  gamma-ray radial distributions predicted by hadronic models are sharper than the
distributions  calculated within the leptonic models (see Fig.13).  However, because of
limited angular  resolution of gamma-ray telescopes,  it is not possible, unfortunately,  to
distinguish   between the distributions predicted by two models. It is demonstrated in Fig.13
in which the profiles are smoothed with a typical  for Cherenkov  telescope arrays  point
spread function assuming Gaussian distribution with $\sigma=0.05^\circ$.  Both
smoothed profiles reasonably agree with  the angular distribution  of TeV gamma-rays as
measured  by HESS.  An improvement of the angular resolution  of future detectors by a
factor of  two  should allow identification of the  origin of the parent particles.
On the other hand,  this can be done with current detectors like HESS, VERITAS and MAGIC,
for a young  nearby SNR  with a larger  angular size. In this regard, the young
supernova remnant  Vela Jr (RX J0852.0-4622)   with gamma-ray
properties quite similar to RX J1713.7-3946 as reported by
HESS (Aharonian et al. \citealp{HESS_VelaJr}), but with angular diameter
$\approx 1^\circ$,   is  a promising object for such studies.

\begin{figure*}[t]
\includegraphics[width=14.0cm]{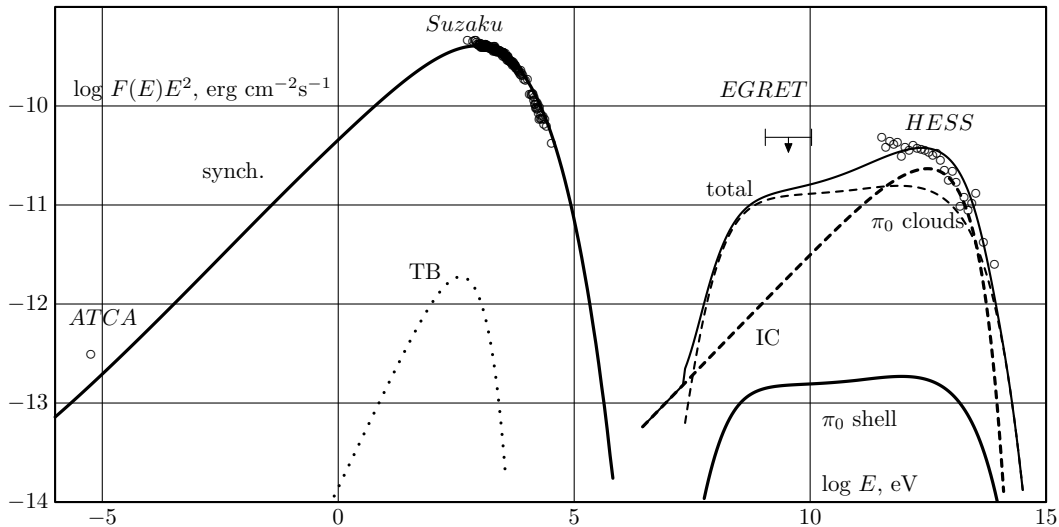}
\caption{Broad-band  emission of RX J1713.7-3946  for the
composite scenario of  gamma-rays with
a non-modified forward shock and dense clouds.  The principal model parameters are:
$t=1620$ yr, $D=1.5$ kpc, $n_H=$0.02 cm$^{-3}$, $E_{SN}=1.2\cdot 10^{51}$ erg,
$M_{ej}=0.74M_{\odot}$, $M^f_A=55$, $M^b_A=10$, $\xi _0=0.1$,
$K^f_{ep}=1.4\cdot 10^{-2}$, $K^b_{ep}=9\cdot 10^{-4}$.
The calculations lead to the following values of the magnetic fields and the shock speeds at
the present epoch:
the magnetic field downstream of the forward and reverse shocks $B_f=22$ $\mu $G and $B_b=31$ $\mu $G,
respectively, the speed of the
forward shock
$V_f=3830$ km s$^{-1}$, the
speed of the reverse shock $V_b=-1220$ km s$^{-1}$. The following
radiation processes are taken into account: synchrotron radiation
of accelerated electrons (solid curve on the left),
thermal bremsstrahlung (dotted line), IC gamma-ray emission of the entire
remnant including forward and reverse shocks (dashed line),
hadronic component of gamma-rays from the remnant's shell (solid line on the right)
as well as from dense clouds assuming the
factor of 120 enhancement of the flux
(thin dashed line). We also show the total gamma-ray emission from the entire remnant including the dense
clouds (thin solid line). }
\end{figure*}

\subsection{Shocked clouds in the shell?}

The observed  ratio of the forward shock radius to the
reverse shock radius robustly   constraints  the type of supernova
explosion that produced SNR RX J1713.7-3946. For reasonable values of the energy of supernova explosion
$E_{SN}<3\cdot 10^{51}$ erg,  the ejecta mass must be small, $M_{ej}<2M_{\odot }$, otherwise the
forward shock speed  would be  too low  for production of  synchrotron radiation extended to
hard X-ray  domain. Such small ejecta masses correspond to Ib/c or IIb core collapse supernova
(Chevalier \citealp{chevalier05}). Circumstellar medium around these types
of supernova is created by the stellar wind of the supernova progenitor. It may be a low density
($n_H\sim 0.01$ cm$^{-3}$) bubble created
by stellar wind of progenitor during the main sequence period of the star evolution or by a stellar wind
of a Wolf-Rayet progenitor.  Because of very low densities,  the observed gamma-rays
can be produced only via IC scattering of electrons. If the  bubble is created during the
Red Supergiant stage of the supernova progenitor or during the interaction of the slow
Red Supergiant wind with  the fast Wolf-Rayet wind, the gas density could be sufficiently high
for effective production of gamma-rays at interactions of accelerated protons.

In all models considered the forward shock of SNR have  swept only
6-15$M_{\odot }$. This mass is comparable or slightly less  than
the mass ejected by progenitors of Ib/c and IIb supernova during
the stellar evolution. This means that the shock just have swept
the progenitor's material ejected during the stellar evolution.
The interaction with the molecular gas surrounding the remnant
only starts. The exceptions could be  the very dense cores of
molecular gas (clouds C and D) that are situated already inside
the forward shock (Fukui \citealp{fukui08}). Probably the forward
shock has enveloped the clouds,  and a high pressure gas from the
downstream region drove  secondary shocks into the clouds (see
Chevalier \citealp{chevalier77} for a review). These shocks are
strongly decelerated and become radiative. That is why they do not
produce the thermal X-rays (see also Cassam-Chena\"i et al.
\citealp{cassam04}). The nonthermal X-rays  can be  produced at
these shocks by high-energy electrons from the remnant's shell in
the compressed magnetic field of the cloud, while the highest
energy gamma-rays from pion decay are mainly produced inside  the
cloud where the target density is high (Fukui \citealp{fukui08}).

If the highest energy protons freely penetrate into the clouds,
the corresponding gamma-emission from the pion decay may exceed
the gamma-emission from the pion decay of the remnant by a factor
that is the ratio of the cloud mass to the mass swept up by the
forward shock. If the clouds C and D indeed contain about 100-1000 solar masses
(Fukui \citealp{fukui08}), this ratio should be  close to
10-100. Since this is not the case, as it follows from the HESS results,
the hadronic models can survive only if the penetration into the
clouds is depressed. Such a possibility  seems unlikely, at least
for the highest energy protons.

On the other hand, in the leptonic models of gamma-rays, in which
the bulk of the gamma-ray emission is produced via IC scattering of electrons,
one may expect  significant contribution of hadronic gamma-rays produced
in the clouds C and D.  Namely the gamma-ray fluxes from p-p interactions
shown in Figs.8\&9  can be enhanced  by 1 or 2 orders of magnitude.
Apparently, from these regions we should expect also
low energy, MeV/GeV  gamma-rays.
Because of low density of the ambient gas, the  gamma-radiation outside the
C and D clouds will be dominated by the IC component, therefore the gamma-ray flux at
MeV/GeV energies is expected to  be  faint (see Figs.8.\&9).

In Fig. 14 we demonstrate the feasibility of the "composite"
scenario. The hadronic component of gamma-rays from dense clouds
is shown by multiplying the flux expected from the low density gas
downstream of the forward shock by a factor  of 120. The ratio of
the gamma-rays from clouds to the flux from other parts of the
shell depends on the ratio of the mass of clouds to the mass of
the shell, provided that all particles freely enter the dense
clouds. This however could not be the case especially for the low
energy particles which introduces large uncertainties in the
predictions of the flux of this component of radiation. On the
other hand, if the enhancement of hadronic gamma-rays in clouds is
indeed very large, this could help to improve the fit of the
gamma-ray data both at low (sub TeV) and high energy (multi-TeV)
regions (see Fig. 14). Note that in this scenario we expect also
gamma-rays from dense clouds due to bremsstrahlung of electrons.
However, as long as the total energy in protons significantly
exceeds the energy in relativistic electrons, the interaction of
latters with the dense gas will dominate over the bremsstrahlung.

\subsection{Reverse shock in hadronic and leptonic models}

The parameters characterizing acceleration of particles in the
reverse shock in the leptonic models are not strongly constrained.
While in the hadronic models the reverse shock must be an efficient accelerator of multi
TeV electrons in order to produce the synchrotron X-radiation
observed from the inner ring,  in the leptonic models the acceleration
by the inverse shock is  not required. Indeed,  in the leptonic models
the electrons accelerated at the forward shock may
easily reach the contact discontinuity downstream of the forward
shock  with relatively weak magnetic fields.
They will produce  X-rays in the amplified
magnetic field between the contact discontinuity and the reverse
shock (see Fig.11). The magnetic field  cannot be weak in this region since the
inner ring is clearly visible in radio. The corresponding
radio-electrons must be accelerated at the reverse shock. The
magnetic field amplified at the reverse shock has probably rather
small scales. This is because the scale of the most unstable MHD
mode is proportional to the strength of the undisturbed field in
the case of the non-resonant streaming instability (Bell
\citealp{bell04}).

The magnetic field of ejecta $B<10^{-8}$G is  significantly
smaller than the field in the circumstellar medium (presumably
about $10^{-6}$ G). The small-scale magnetic field is not a very
efficient scatter of highest energy electrons. As a result,  the
diffusion may be significantly faster than the Bohm diffusion, and
the acceleration at the reverse shock up to TeV energies might be
impossible. However,  acceleration of GeV electrons producing
synchrotron radio emission is quite probable. On the other hand,
if the reverse shock is an effective accelerator,  it may produce
the additional component(s)  of nonthermal radiation. For example,
in the case of  weak magnetic fields  at the reverse shock,  the
hadronic models may result in  detectable IC gamma-rays from the
central region of the remnant.  Probably the situation is
different in the remnant Cas A, where the magnetic field at the
reverse shock is stronger than the magnetic field at the forward
shock. This may be related with a different acceleration
efficiency of the shocks in the remnant Cas A. For example,  the
reverse shock may be modified and the forward shock not. The
reverse shock energetics in Cas A is of the order of 10 \% from
the total one. It is enough to explain the observable TeV
gamma-rays in the framework of hadronic models.

\begin{figure}[t]
\includegraphics[width=8.0cm]{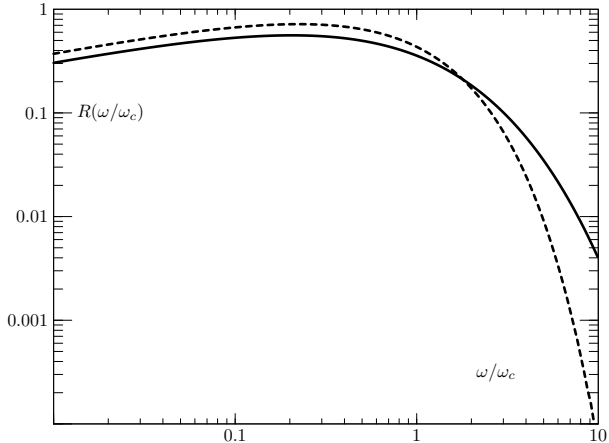}
\caption{ The synchrotron emissivity of a single electron in the
isotropic magnetic field. The function $R(x)$ for single value of
the magnetic field strength given by Eq.(A2)
(dashed line) and the function $R_1(x)$ (solid
line)  given by Eq. (A5) are shown. }
\end{figure}

\begin{figure}[t]
\includegraphics[width=7.5cm]{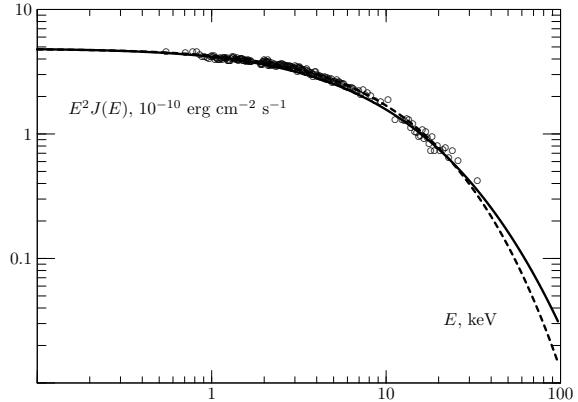}
\caption{ Spectra of synchrotron X-rays produced at the  shock with a distributed magnetic field
(Eq. (B5), solid line) and at the
 shock with a single-valued magnetic field (Eq.(B3), dashed line).
Suzaku data  (circles) for the remnant RX J1713.7-3946 are also shown. }
\end{figure}

\begin{figure}[t]
\includegraphics[width=7.5cm]{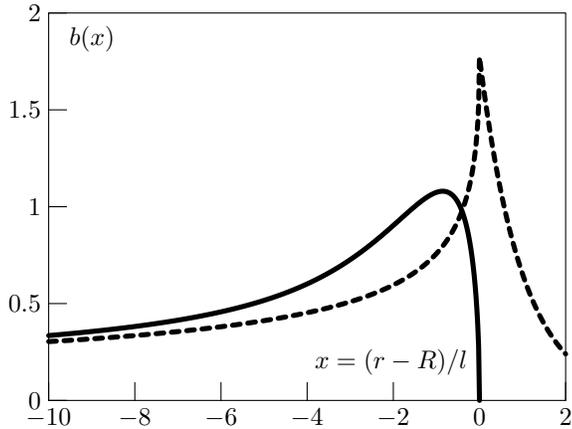}
\caption{The function $b(x)$ that describes the radial profile of the
X-ray brightness for the forward shock (Eq. (D2), solid line) and for the
reverse shock (Eq.(D3), dashed line).}
\end{figure}

\section{Conclusion}

Our study of  the nonlinear diffusive shock acceleration in RX ~J1713.7-3946, and
calculations of  the related electromagnetic radiation lead to  the following conclusions:

1. The present reverse shock position may be reproduced if
the SNR RX J1713.7-3946 is the remnant of Ib/c or IIb type supernova with
the ejected mass $M_{ej}<2M_{\odot }$.

2.  The available radio, X-ray and gamma-ray data can be explained by  both the
leptonic and  hadronic models, although with certain caveats. The spectral
shape of gamma-emission is better reproduced in  hadronic models.  The morphology
of electromagnetic emission in X-ray and TeV gamma-ray bands
can be  in principle reproduced  by  both the   hadronic and leptonic models.

3. The hadronic origin is possible only if the following conditions are satisfied:\\
a) The forward shock is strongly modified throughout  all surface.\\
b) Chemical abundance of heavy ions in  X-ray emitting regions  should be significantly
below the solar one. \\
c)  The lack of the predicted  by hadronic models sharp outer X-ray rim
can be explained by the  nonuniform circumstellar medium.\\
d) If the forward shock interacts with clouds D and C,  the
penetration of the highest energy protons into the clouds must be
suppressed.

4.   Any detection of  GeV gamma-rays by
AGILE and Fermi  satellites at the level of $F \geq  10^{-11}$ erg cm$^{-2}$ s$^{-1}$, especially
outside the dense clouds C and D,  would be a decisive argument in favor of
the hadronic origin of TeV gamma-rays. The  leptonic models predict very hard energy
spectra of gamma-rays with  absolute energy fluxes  as low as $F=2\cdot 10^{-12}$ erg cm$^{-2}$ s$^{-1}$.
However a  low flux and hard gamma-ray spectrum  are expected also  in the
hadronic models  with dense filaments, assuming energy-dependent cosmic ray penetration into the clouds.

5.  The leptonic models imply  that the bulk of the observed gamma-radiation is produced via IC scattering
of electrons. However, these models  allow also significant contributions of hadronic gamma-rays
from  dense gas region in the shell, in particular from the clouds C and D. In such "composite" models
we should  expect   relatively high fluxes of low energy gamma-rays from the C and D clouds, while outside these
condensations the MeV/GeV flux is expected to be quite low.

6. If the inner ring of RX J1713.7-3946 is indeed produced by the reverse shock,  its proper motion
can be measured. X-ray data of Chandra may be used for this purpose since the X-ray filaments in the inner
ring region have been  observed (Lazendic et al. \citealp{lazendic04}, Uchiyama et al. \citealp{uchiyama03}).
In all models considered
the reverse shock moves with several thousand kilometers per second to the center of the remnant. The model
with $r^2$ density profile has the highest absolute value of  the reverse shock speed,
$V_b=-3010$ km s$^{-1}$ or even larger for a sharper density profile.  The measurement of the
reverse shock speed  will constrain the  gas density distribution around this SNR.

\begin{acknowledgements}
We thank T. Tanaka for providing us with the Suzaku X-ray data.
We are grateful to H. V\"olk for the useful discussions
on cosmic ray acceleration in SNRs.
VNZ acknowledge the hospitality of the Max-Planck-Institut f\"ur
Kernphysik, where the major part of this work was carried out.
The work of VNZ was also supported by the RFBR grant in Troitsk.

\end{acknowledgements}

\appendix

\section{Synchrotron emission in the turbulent medium}

The synchrotron emissivity in the medium with nonuniform magnetic
field can be written as

\begin{equation}
\epsilon (\omega )=\frac {\sqrt{3}q^3}{2\pi m_ec^2}
\int dB BP(B)\int p^2dpN(p)R(\omega /\omega _c)
\end{equation}
where $\omega _c$ is the characteristic frequency of synchrotron
radiation $\omega _c=1.5\ qBp^2/m_e^3c^3$. The function $P(B)$ is the
probability distribution of the magnetic field. If the magnetic field
strength has a single value $B_0$, then $P(B)=\delta (B-B_0)$ and
this equation is reduced to a standard formula for the synchrotron emissivity.

The function $R(x)$ of the argument $x=\omega /\omega _c $
describes synchrotron radiation of a single electron in magnetic
field with chaotic directions.  Crusius \& Schlickeiser
\cite{crusius86} derived an exact expression for  $R(\omega
/\omega _c)$ in terms of Whittaker's function. With an accuracy of
several percent $R(\omega /\omega _c)$ can be presented in a
simple analytical form (Zirakashvili \& Aharonian \citealp{zirakashvili07b}):
\begin{equation}
R(x)= \frac {1.81\exp (-x)}
{\sqrt{x^{-2/3}+(3.62/\pi )^2}} \ .
\end{equation}

In the turbulent medium the magnetic field is not uniform and
is described by a probability distribution $P(B)$. It reveals the  exponential tails when the
magnetic field is amplified by turbulent motions (see Schekochihin et al. \citealp{schekochihin04}).
The probability distribution found in the numerical  simulations of the nonresonant streaming instability
may be written in the simple analytical form (Zirakashvili \& Ptuskin \citealp{zirakashvili08}):

\begin{equation}
P(B)=\frac {6B}{B_{rms}^2}\exp (-\sqrt{6}B/B_{rms}) .
\end{equation}

Here $B_{rms}=\left< B^2\right> ^{1/2}$ is the square root of the mean square of the random magnetic field
(note the corrected missprint in the normalization factor in this equation).
Now the synchrotron emission of a single electron is described by the function $R_1(x)$:

\begin{equation}
R_1(x)=6\int \limits ^{\infty }_{0}dyy^2R(x/y)\exp (-\sqrt{6}y) .
\end{equation}

We found a useful analytical approximation of this function

\begin{equation}
R_1(x)=1.50x^{1/3}\left(1+1.53x^{1/2} \right) ^{11/6}\exp \left( -96^{1/4}x^{1/2}\right)
\end{equation}

The functions $R(x)$ and $R_1(x)$ are shown in Fig. (15).

Qualitatively similar results were found by Bykov et al.
\citealp{bykov08}, but assuming Gaussian
probability distribution for $P(B)$.

\section{Analitical expressions for the synchrotron X-rays}

The analytical approximations for the electron and synchrotron
spectra at the non-modified shock were found by Zirakashvili \&
Aharonian \citealp{zirakashvili07}. The synchrotron loss dominated
regime  was considered. In the case of the single valued magnetic
field the synchrotron spectra are parameterized with  the principal parameter
$E _0$:

\begin{equation}
E _0=\frac {81}{64(1+\kappa ^{1/2})^2}\frac {\hbar m_ecV_f^2}{q^2\eta _B}=
\frac {\mathrm{2.2\ keV}}{(1+\kappa ^{1/2})^2\eta _B}
\left( \frac {V_f}{3\cdot 10^3 \ \mathrm{km\ s}^{-1}}\right) ^2 .
\end{equation}
Here $\kappa $ is the ratio of the upstream and downstream magnetic field strength.
In the case when magnetic field has the same value upstream and downstream of the shock ($\kappa =1$) the
spectrum of synchrotron X-rays has the following form

\begin{equation}
F(E)\propto E^{-2}\left[ 1+0.46\left( \frac {E}{E_0}\right) ^{0.6}\right] ^{11/4.8}\exp
\left( -\sqrt{\frac {E}{E_0}}\right)
\end{equation}

In the case when ratio of the upstream and downstream magnetic field strength
$\kappa =1/\sqrt{11}$, the  spectrum of synchrotron X-rays is mainly determined
by the downstream region and has the following form

\begin{equation}
F(E)\propto E^{-2}\left[ 1+0.38\left( \frac {E}{E_0}\right) ^{0.5}\right] ^{11/4}\exp
\left( -\sqrt{\frac {E}{E_0}}\right) .
\end{equation}
Suzaku data obtained for RX J1713.7-3946 in a broad energy interval from 0.4 to 40 keV
are well fitted by these spectra (Uchiyama et al. \citealp{uchiyama07},
Tanaka et al. \citealp{tanaka08}). For example,  $E_0\approx 0.94$ keV for  Eq.(B3).

In the case of distributed magnetic field one should use  Eq.(A5) instead of Eq.(A2).
Using the electron spectra of Zirakashvili \& Aharonian \citealp{zirakashvili07} we found the following
synchrotron spectra

\begin{equation}
F(E)\propto E^{-2}\left[ 1+0.172\left( \frac {E}{E_1}\right) ^{0.46}\right] ^{25/(12\cdot 0.46)}
\exp \left[ -\left( \frac {E}{E_1}\right) ^{1/3}\right] , \ \kappa =1,
\end{equation}
and
\begin{equation}
F(E)\propto E^{-2}\left[ 1+0.185\left( \frac {E}{E_1}\right) ^{0.4}\right] ^{25/(12\cdot 0.4)}
\exp \left[ -\left( \frac {E}{E_1}\right) ^{1/3}\right] , \ \kappa =1/\sqrt{11}.
\end{equation}
Here the parameter $E_1$ is

\begin{equation}
E _1=\frac {\sqrt{6}}{128(1+\kappa ^{1/2})^2}\frac {\hbar m_ecV_f^2}{q^2\eta _B}=
\frac {\mathrm{0.134\ keV}}{(1+\kappa ^{1/2})^2\eta _B}
\left( \frac {V_f}{3\cdot 10^3 \ \mathrm{km\ s}^{-1}}\right) ^2 .
\end{equation}
Suzaku data may be fitted with $E_1\approx 0.036$ keV using the formula (B5). This value
and $\eta _B=2$ corresponds to the shock velocity $V_f=3300$ km s$^{-1}$. This number is slightly
above the corresponding values in the hadronic scenario when the larger compression ratio of the
modified shocks and the additional input of the faster reverse shock should be taken into account.
The shock velocity is higher in the leptonic scenario when the adiabatic losses
downstream of the shock play a role.

The spectra given by Eqs.(B3),(B5) and Suzaku data are shown in Fig.16.
The distributed magnetic field makes the cut-off of synchrotron spectra slightly smoother in comparison with
a single-value case. The cut-off energy is a factor of 1.6 higher for the distributed magnetic field.

The spectrum (B5) shows a rather slow drop beyond the cut-off. It may be used for explanation of the high-energy
X-ray tail observed in the supernova remnant  Cas A.

\section{Simple estimates of the gamma-ray emission and thermal X-ray emission in SNR. }

Using Eq. (10) one can estimate the characteristic electron temperature downstream of the forward shock.
Assuming a linear radial profile of the gas velocity we found the following expression for the  maximum
electron temperature

\begin{equation}
T_e=\left( \xi _g(n_H+4n_{He})\lambda _{ep}RV_f\frac {4\sigma \sqrt{2m_e}q^{4}}{3(\sigma -1)m}\right) ^{2/5}
=\ 1.6\mathrm {keV}
\ n_H^{2/5}\left( \frac {\sigma \xi _g}{\sigma -1}\right) ^{2/5}
\left( \frac {R_f}{3 \ \mathrm{pc}}\right) ^{2/5}
\left( \frac {V_f}{3\cdot 10^3 \ \mathrm{km\ s}^{-1}}\right) ^{2/5}
\end{equation}
Here $\xi _g=P_g/\rho _0V_f^2$ is the ratio of the gas pressure downstream the shock $P_g$ to the
dynamical pressure $\rho _0V_f^2$ and $\sigma $ is the total compression ratio of the shock. The time
derivative of the temperature was neglected in Eq.(10).
For cosmic ray modified shock in the hadronic model $\xi _g\sim 0.25$, $\sigma =6.5$
and we obtain $T_e=0.6$ keV in agreement with the numerical results.

It is clear from Eq.(C1) that the dependence of the electron temperature on the shock parameters is rather
weak. E.g at the Sedov stage $T_e\propto t^{-0.08}$. It also weakly (only a factor of 2)
affected by the shock modification.
Note that the electron-ion equilibration case when the electron temperature $T_e\propto V_f^2$ and may
be strongly reduced by the shock modification (see e.g. Drury et al. \citealp{drury09}).

X-ray emissivity due to the thermal bremsstrahlung is given by (Rybicki \& Lightman \citealp{rybicki79})

\begin{equation}
E^2\epsilon _{ff}=\frac {8q^6}{3m_ec^3h}
\left( \frac {2\pi }{3m_eT_e}\right) ^{1/2}E\sum \limits _iZ^2n_en_i\tilde{g}_{ff}
\exp \left( -\frac {E}{T_e}\right) .
\end{equation}
Here $Z$ is the charge number of plasma ions, $\tilde{g}\sim 1$ is the Gount factor. Assuming the radial density
profile downstream of the shock $\rho \propto r^{3(\sigma -1)}$ and the uniform electron temperature we find
the flux of the thermal X-rays

\[
E^2F_X=\frac {\sigma ^2}{6\sigma -3}
\frac {8q^6}{3m_ec^3h}\left( \frac {2\pi }{3m_eT_e}\right) ^{1/2}\left( 1+2\frac {n_{He}}{n_H}\right)
\left( 1+4\frac {n_{He}}{n_H}\right) \frac {n_H^2R_f^3}{D^{2}}
\frac {E}{T_e}\exp \left( -\frac {E}{T_e}\right) \tilde{g}_{ff}=
\]
\begin{equation}
2.5\cdot 10^{-11}\frac {\mathrm{erg}}{\mathrm{s\ cm}^2}\frac {\sigma ^2}{6\sigma -3}
n_H^2R^3_{pc}D^{-2}_{kpc}T^{1/2}_{keV}
\frac {E}{T_e}\exp \left( -\frac {E}{T_e}\right) .
\end{equation}

The gamma-ray flux from the pion decay at energies below the cut-off may be estimated as
(e.g. Zirakashvili \citealp{zirakashvili08b})

\begin{equation}
E^2F_{pp}(E)=\frac {R_f^3K_{\pi \pi }\sigma _{pp}cn_H^2\xi_{cr}mV_f^2}{D^2\ln (p_{\max }/mc)}
\left( 1+4\frac {n_{He}}{n_H}\right) ^2=
1.6\cdot 10^{-11}\frac {\mathrm{erg}}{\mathrm{s\ cm}^2}\xi _{cr}R^3_{pc}D^{-2}_{kpc}n_H^2
\left( \frac {V_f}{3\cdot 10^3 \ \mathrm{km\ s}^{-1}}\right) ^2
\end{equation}
Here $K_{\pi \pi }=0.17$ is the fraction of the proton energy
transmitted to the parent neutral pions, $\sigma _{pp}$ is the
total inelastic $p-p$ cross-section, $p_{\max }$ is the maximum
momentum of accelerated protons and  $\xi _{cr}$ is
 the ratio of the cosmic ray pressure downstream of the shock $P_{cr}$ to the
dynamical pressure $\rho _0V_f^2$. We use the value
$\sigma _{pp}=4\cdot 10^{-26}$ for this estimate of the flux of TeV gamma-rays. It was assumed that
the accelerated protons with power-low $E^{-2}$ distribution fill the remnant uniformly.

The intensity of gamma-rays from the IC scattering in the synchrotron losses dominated case may be estimated as
(Zirakashvili \citealp{zirakashvili08})

\begin{equation}
E^2F_{IC}=\frac {3\xi _{cr}}{8\xi _B}\frac {K_{ep}R_f^2U_{rad}V_f}{D^2\ln (p_{\max }/mc)}=
9\cdot 10^{-12}\frac {\mathrm{erg}}{\mathrm{s\ cm}^2}R^2_{\theta }(deg)
\frac {V_f}{3\cdot 10^3 \ \mathrm{km\ s}^{-1}}
\frac {K_{ep}}{10^{-2}}\frac {\xi _{cr}}{\xi _B}
\frac {U_{rad}}{4\cdot 10^{-13}\mathrm{erg\ cm}^{-3}}
\end{equation}
Here $\xi _B$ is the ratio of the magnetic energy $B^2/8\pi $ to the dynamical pressure $\rho _0V_f^2$, $U_{rad}$
is the energy density of the scattered photons and $R_{\theta }(deg)$ is the angular radius of the shock in
degrees.

The ratio of the  gamma-ray energy flux to the thermal bremsstrahlung energy flux at
its maximum at $E_X=T_e$ is given by equation
(see also Katz \& Waxman \citealp{katz08})
\begin{equation}
\frac {E^2F_{pp}}{E^2_XF_X}=1.7\xi _{cr}\frac {6\sigma -3}{\sigma ^2T^{1/2}_{keV}}
\left( \frac {V_f}{3\cdot 10^3 \ \mathrm{km\ s}^{-1}}\right) ^2
\end{equation}
These fluxes are comparable for cosmic-ray modified shocks with $\xi _{cr}\sim 0.5$ and $\sigma \sim 6$.

\section{Projection effect for a forward and a reverse shock}

The surface brightness $J$ is determined by the integral of emissivity $\epsilon $ along the line of sight.
For spherically
symmetric case it can be written as

\begin{equation}
J(r)=2\int \limits _{r}^{\infty }\frac {r'dr'\epsilon (r')}{\sqrt{r'^2-r^2}}
\end{equation}

For a forward shock the emissivity downstream can be written as $\epsilon =\epsilon _0H(R-r)\exp ((r-R)/l)$.
Here $H(r)$ is the step function. For a reverse shock the emissivity has the form
$\epsilon =\epsilon _0H(r-R)\exp ((R-r)/l)$. For a thin shell of X-ray emission $l<<R$ the surface
brightness $J(r)=\sqrt{2Rl}\epsilon _0b(x)$ may be written in terms of the function $b(x)$ of argument $x=(r-R)/l$.

For the forward shock this function (see also Ballet \citealp{ballet06})

\begin{equation}
b(x)=
\left\{ \begin{array}{ll}
2\exp (x)\int \limits _{0}^{\sqrt{-x}}dy\exp (y^2), \ x<0 \\
0, \ x>0
\end{array} \right.
\end{equation}

while for the reverse shock
\begin{equation}
b(x)=\exp (-x)
\left\{ \begin{array}{ll}
2\int \limits _{\sqrt{-x}}^{\infty }dy\exp (-y^2), \ x<0  \\
\sqrt{\pi }, \ x>0
\end{array} \right.
\end{equation}

The function $b(x)$ is shown in Fig.17. For the forward shock the function $b(x)$ has a width
$w=4.6l$ at the half of maximum and $w=7.5l$ at the $1/e$ of maximum. These numbers are in agreement
with results of Ballet \citealp{ballet06} and Berezhko \& V\"olk \citealp{berezhko04a}. For the
reverse shock this function has  a  width $w=1.3l$ at the half of maximum and $w=2.6l$ at the $1/e$
of maximum.

%

\end{document}